\documentclass[preprint]{aastex}
\usepackage{color}
\usepackage{natbib}
\usepackage{multirow}

\begin{document}

\title{ALMA and IRIS Observations of the Solar Chromosphere II: Structure and Dynamics of Chromospheric Plage}

\author{Georgios Chintzoglou\altaffilmark{1,2}, Bart De Pontieu\altaffilmark{1,4,5}, Juan Mart\'inez-Sykora\altaffilmark{1,3,4}, Viggo Hansteen\altaffilmark{1,3,4,5}, Jaime de la Cruz Rodr\'iguez\altaffilmark{6}, Mikolaj Szydlarski\altaffilmark{4,5}, Shahin Jafarzadeh\altaffilmark{4,5}, Sven Wedemeyer\altaffilmark{4,5}, Timothy S. Bastian\altaffilmark{7} and Alberto Sainz Dalda\altaffilmark{1,3,8}}

\email{gchintzo@lmsal.com}

\altaffiltext{1}{Lockheed Martin Solar \& Astrophysics Laboratory, Palo Alto, CA 94304, USA}
\altaffiltext{2}{University Corporation for Atmospheric Research, Boulder, CO 80307-3000, USA}
\altaffiltext{3}{Bay Area Environmental Research Institute, NASA Research Park, Moffett Field, CA 94035, USA}
\altaffiltext{4}{Rosseland Center for Solar Physics, University of Oslo, P.O. Box 1029 Blindern, NO0315, Oslo, Norway}
\altaffiltext{5}{Institute of Theoretical Astrophysics, University of Oslo, P.O. Box 1029 Blindern, NO0315, Oslo, Norway}
\altaffiltext{6}{Institute for Solar Physics, Department of Astronomy, Stockholm University, AlbaNova University Centre, SE-106 91, Stockholm, Sweden}
\altaffiltext{7}{National Radio Astronomy Observatory, 520 Edgemont Road, Charlottesville, VA 22903, USA}
\altaffiltext{8}{Stanford University, HEPL, 466 Via Ortega, Stanford, CA 94305-4085}

\begin{abstract}
We propose and employ a novel empirical method for determining chromospheric plage regions, which seems to better isolate plage from its surrounding regions compared to other methods commonly used. We caution that isolating plage from its immediate surroundings must be done with care in order to successfully mitigate statistical biases that, for instance, can impact quantitative comparisons between different chromospheric observables. Using this methodology, our analysis suggests that $\lambda$=1.25\,mm free-free emission in plage regions observed with \emph{ALMA}/Band6 may not form in the low chromosphere as previously thought, but rather in the upper chromospheric parts of dynamic plage features (such as spicules and other bright structures), i.e., near geometric heights of transition region temperatures. We investigate the high degree of similarity between chromospheric plage features observed in \emph{ALMA}/Band6 (at 1.25 mm wavelength) and \emph{IRIS}/\ion{Si}{4} at 1393\,\AA. We also show that \emph{IRIS}/\ion{Mg}{2} h and k is not as well correlated with \emph{ALMA}/Band6 as was previously thought, and we discuss the discrepancies with previous works. Lastly, we report indications for chromospheric heating due to propagating shocks supported by the \emph{ALMA}/Band6 observations.

\end{abstract}


\section{Introduction}\label{intro}

The chromosphere is the most complex and remains one of the least understood layers of the solar atmosphere. It is the layer where the atmosphere transitions from a plasma-dominated to a magnetic-field-dominated regime. It is a medium where ion-neutral interactions matter (such as ambipolar diffusion), and a place in the atmosphere where radiative transfer effects from departures from local thermodynamic equilibrium (non-LTE) are important. In addition, the chromosphere is the atmospheric shell through which energy and mass from the photosphere must pass in order to heat the overlying corona and power the solar wind.

From a modern (yet historical) observational standpoint, the chromosphere can be split into three general regions: i.e., (a) active region (AR), sunspots and surroundings, (b) quiet regions, and (c) regions of ``\emph{plage}''. The AR chromosphere plays a dominant role in the energy release during intense solar flares. Often, filaments are seen to form and to erupt from neutral lines produced by flux emergence or decay (e.g., \citealp{Chintzoglou_etal_2017, Chintzoglou_etal_2019}). Sunspots seen in chromospheric wavelengths (e.g., in H$\alpha$) exhibit ``superpenumbras'', i.e., a system of chromospheric fibril-like structures lying above the penumbra and which are oriented radially outwards, typically extending beyond the end of the penumbra. Superpenumbras comprise a distinct region near sunspots, commonly observed in the chromosphere of well-developed sunspots. The superpenumbral fibrils suggest, in general, a more horizontal magnetic orientation in the chromosphere, i.e., a canopy-like topology. Several studies (e.g., \citealt{Yurchyshyn_etal_2001a, Zhang_Jun_etal_2003}) have explored the association of superpenumbras with moving magnetic features (MMFs; \citealt{Harvey_Harvey_1973}) rushing radially outwards from sunspots in the photosphere.

A major aspect of the physics of the quiet chromosphere is the ubiquitous presence of shocks, propagating from the photosphere upwards \citep{Carlsson_Stein_1997}. Such shocks may play an important role in energizing the quiet chromosphere. However, the most conspiquous features of the quiet chromosphere are the so-called spicules \citep{Secchi_1877}. Spicules are jets of chromospheric material seen as rooted at the chromospheric network. About a decade ago, a new class of spicules, known as ``Type-II spicules'', was found in high-resolution imaging observations taken at the \ion{Ca}{2} H line \citep{DePontieu_etal_2007b}. These structures are more slender (apparent widths $\approx 1\arcsec$) and exhibit higher plane-of-the-sky speeds ($\approx 50 - 100$ km s$^{-1}$) than their ``traditional'' counterparts.

Chromospheric \emph{plages} are regions of higher intensity in chromospheric lines (e.g., traditionally in \ion{Ca}{2} H\&K, or H$\alpha$) with stronger magnetic fields as compared to the typical quiet Sun, but weaker if compared to those in a sunspot chromosphere. Therefore, plages can be viewed as areas in the solar chromosphere that are intermediate between sunspots and the quiet Sun conditions. In such plage regions we find short and dynamic structures called ``dynamic fibrils'', which are driven by slow-mode magnetoacoustic shocks that propagate from the photosphere to the chromosphere and beyond \citep{Hansteen_etal_2006, DePontieu_etal_2007a, Skogsrud_etal_2016, Carlsson_etal_2019}. 

The definition of plage has never been done strictly; it was -- and remains -- a loose observational term. \citet{Deslandres_1893} first used the term ``plage'' metaphorically (which in French means ``the beach'' or ``seashore''), to help explain his interpretation that such bright regions seen in his early \ion{Ca}{2} H and K full-disk spectroheliograms correspond to elevated structures sticking out of the photosphere, in an analogy to sandy beaches emerging from (and appearing brighter than) the ocean\footnotemark. As a result of this loosely-defined term that so poetically describes regions with higher intensity of chromospheric lines as compared to those in the typical quiet Sun, early and modern observers have been determining such areas on the Sun with a relative freedom; either (a) ``by eye'', i.e., manual definition of the boundaries of a plage region (for a recent example see, e.g., \citealt{Carlsson_etal_2015}); or, (b) in a quantitative manner, i.e., by selecting a certain intensity threshold for a chromospheric image, or even an area of moderately intense magnetic fields (e.g., \citealt{Jafarzadeh_etal_2019} and references therein). With the present paper we raise caution in that the exact method of identifying and isolating plage from its immediate surroundings (e.g., sunspots, pores, quiet Sun) can introduce statistical biases which can have a significant impact on quantitative comparisons between chromospheric observables (see \S~\ref{plagemethod}, \S~\ref{comparison}, \S~\ref{discrepancies}, and APPENDIX).

\footnotetext{As it appears, Deslandres introduced the term \emph{plage} to Solar Physics rather unintentionally, because after a full paragraph of using that term (as a metaphor) he concluded that these bright chromospheric regions shall be named ``\emph{flammes faculaires}'' (i.e., ``facular flames'', due to the association with faculae in the photosphere; \citealt{Deslandres_1893}).}

Quantitative comparisons between different observed diagnostics are a key way to study the physics of the chromospheric plasmas and to understand the diagnostic power of various chromospheric observables. Recently, the \emph{Atacama Large Millimeter/submillimeter Array} (\emph{ALMA}; \citealt{Wootten_Thompson_2009}) has offered its unique capability in producing high-resolution ($<1\arcsec$) and fast time-cadence (2 sec) imaging of free-free emission (from chromospheric electrons) in the millimetric (mm) part of the spectrum. 

\bf Under chromospheric conditions the source function, $S_\lambda$, of the free-free emission at mm-wavelengths is in LTE and so the source function is the Planckian, $S_\lambda=B_\lambda(T)$. \rm In addition, since the radiation is in mm-wavelengths (i.e., low frequencies) the Rayleigh-Jeans approximation holds true, which dictates that the source function is linearly proportional to the blackbody temperature. This mm-emission becomes optically thick over a rather narrow width of heights. Given that, we can measure the ``brightness temperature'', $T_b$, i.e., a temperature a blackbody would have to match the brightness of the observed emission. $T_b$ can thus be used to infer the local plasma temperature. 
The local conditions producing the optically-thick free-free emission can originate from quite a wide range of geometric heights. Additionally, the formation height depends on the electron density, which is also expected to vary wildly in the chromosphere. Conversely, we do not know the exact height where the free-free emission becomes optically thick and therefore, we do not know where exactly $T_b$  is measured \citep{Carlsson_Stein_2002, Wedemeyer_etal_2007, Loukitcheva_etal_2015, Rutten_2017, Martinez-Sykora_etal_2020a}. \citet{Rutten_2017} argued that \emph{ALMA} mm-emission should be dominated by fibrils and spicules along the canopy as typically seen in H$\alpha$ or with even higher opacities. In the present paper we are taking advantage of the high spatial resolution \emph{ALMA} observations at $\lambda$=1.25\,mm and address the formation height problem for this free-free mm-emission. 

\citet{Bastian_etal_2017} presented the first quantitative comparison between UV chromospheric emission and \emph{ALMA} free-free emission. These authors explored how well the $T_b$ measured with \emph{ALMA} at 1.25\,mm correlate with the average chromospheric radiative temperatures, $T_{\rm rad}$, inferred by converting the average \emph{IRIS}/\ion{Mg}{2} h2v and h2r peak intensities, $I_\lambda$, into $T_{\rm rad}$ via the Planck function: 

\begin{equation}\label{planck}
T_{\rm rad}=\frac{hc}{k_B\lambda}\frac{1}{ln(\frac{2hc^2}{\lambda^5I_\lambda}+1)}
\end{equation}

\noindent where, $c$, the speed of light, $h$, the Planck's constant, and, $\lambda$, the average wavelength position of \ion{Mg}{2} h2v and h2r peaks. This study reported a positive correlation with some scatter mainly attributed to the expectation that $T_{\rm rad}$ might not be perfectly correlated with $T_b$, since the source function for \ion{Mg}{2} h2v/h2r decouples from the local temperature with increasing height in the atmosphere (because \ion{Mg}{2} k\&h are scattering lines). Similar results were found by \citet{Jafarzadeh_etal_2019} using the same \emph{ALMA}/\emph{IRIS} observations as in \citet{Bastian_etal_2018} although they studied correlations between \emph{ALMA}/Band6 $T_b$ with $T_{\rm rad}$ from \ion{Mg}{2} but for each of its h and k line features individually. \citet{DaSilvaSantos_etal_2019} performed inversions of \emph{IRIS} observations and used \emph{ALMA} data as an additional constraint. Apart from several low temperature regions they also found high temperature regions which seem to be associated with shocks pervading the chromosphere. \citet{Wedemeyer_etal_2020} presented \emph{ALMA}/Band3 interferometric maps and discussed the potential of such observations for the study of the dynamic chromosphere on small scales (such as small loops).

In this paper, we composed and analyzed a unique and comprehensive dataset from joint observations with \emph{ALMA}, the \emph{Interface Region Imaging Spectrograph} (\emph{IRIS}; \citealt{DePontieu_etal_2014}), and the \emph{Solar Dynamics Observatory} (\emph{SDO}; \citealt{Pesnell_etal_2012}). Our dataset is most appropriate for investigating the rich dynamics of the solar chromosphere and transition region in plage and its peripheral areas -- including spicules and chromospheric shocks -- thanks to the synergy of high spatial and temporal resolution of spectral and imaging observations by \emph{IRIS} with fast time-cadence and unique temperature diagnostic capabilities from \emph{ALMA} interferometric observations. \bf A companion publication (Paper I; Chintzoglou et al 2020a) focuses on the evolution of a spicule in the western part of the \emph{IRIS} raster. \rm Here, using the same dataset, we focus on the general structure and the dynamics of chromospheric plage. Since some of our results show discrepancies with those reported by several previous studies, we perform and present a thorough comparison to elucidate the reasons behind the discrepancies. For carrying this investigation, we introduced a novel empirical methodology to better determine the boundaries of regions of plage in the observations and we also employed a state-of-the-art numerical model to synthesize observables for comparison with the observations.

This paper is structured as follows: in \S~\ref{obs} we provide a description of the observations and the model used in this work. In \S~\ref{plagemethod} we describe our proposed methodology for the determination of pure regions of plage, and in \S~\ref{results} we continue with our analysis and presentation of our results, followed by a discussion of the discrepancies in \S~\ref{discrepancies}. We close with our summary and conclusions in \S~\ref{conclusion}.

\section{Observations and Modeling}\label{obs}


We observed a plage region in the leading part of NOAA AR12651 on 22-Apr-2017, at heliographic coordinates N11$\degr$E17$\degr$, or at ($x$,$y$)=(-260$\arcsec$, 265$\arcsec$) in helioprojective coordinates (Figure~1a Paper I). The overall spatial distribution of the plage fields in the target appeared semicircular in shape, as organized around the outer boundary of a supergranule (Figure~1b Paper I). The common \emph{IRIS}-\emph{ALMA} field of view (FOV) contained part of that plage, including a photospheric pore (e.g., Figure~1c Paper I). A very high degree of similarity between morphological structures seen in \emph{ALMA}/Band6 maps and \emph{IRIS}/SJI 1400\,\AA\ images was evident in our observations (Figure~1 of Paper I). We address the origin of this outstanding similarity in \S~\ref{RESULTS_SPICULE}. \bf For additional details regarding the reduction of the \emph{IRIS} and \emph{ALMA} observations used here refer to \S~2 of Paper I. \rm
\emph{ALMA} captures in ultra-high cadence (2\,s) dynamics in plage and interesting evolution of linear-like structures, including indications for shocks in the region of plage. In this work, we address the nature of the high correlation we found between spatially-resolved features seen in \emph{IRIS}/\ion{Si}{4} and \emph{ALMA}/Band6 and in other observables both in the observations and in the model (\S~\ref{RESULTS_SPICULE}). 

\subsection{General Morphology of the Dynamic Plage in \emph{IRIS} and \emph{ALMA}/Band6 Observations}

 To a top-level view, the chromospheric intensities in the observables (Figure~\ref{FIG_2}) are brightest in the strongly magnetized regions and dimmer in the interloping weakly magnetized area. Our \emph{ALMA}/\emph{IRIS} raster FOV contains plage in the north and south parts but also contains the outskirts of plage, i.e., plage ``periphery''. \bf For context imaging and for an image of the overal magnetic distrinution in our target region see Figure 1b of Paper I. \rm Here, we address the physics of plage and the problem of defining regions of plage. Therefore, we study the full FOV as well as the individual parts in the FOV that are designated as (a) plage and (b) plage periphery (see following sections). Last, in the same Figure~\ref{FIG_2} we indicate locations of intensity features in \ion{SI}{4} 1392\,\AA\ rasters that show outstanding similarities (intensity correlations) with \emph{ALMA}/Band6 (boxed regions). Remarkably, these regions appear to be weakly correlated, or even found in anti-correlation between \ion{Mg}{2} 2796\,\AA\ rasters and \emph{ALMA}/Band6. We address the origin of this finding with a rigorous analysis of the observations and we discuss the physical implications using an advanced MHD model in \S~\ref{results}.

\subsection{Bifrost Simulation of Dynamic Plage and Synthesis of \emph{ALMA} and \emph{IRIS} Observables}\label{observables}

The simulation analyzed in this paper produces Type-II spicules in several locations in the computational domain, in-between regions of emerging flux and plage (the latter containing dynamic fibrils). Additional details regarding the simulation and the synthesis of observables can be found in Paper I and in \citet{Martinez-Sykora_etal_2020b}. 

In order to perform a comparison of the physics and the evolution of the observed plage and its periphery with those in the simulation, we focus on two particular regions, (1) a low magnetic flux ``spicule region'' at $x=[40,45]$\,Mm (seen to develop at a favorable angle with the line of sight (LOS), Figure~\ref{FIG_4}ab, annotated and pointed with arrows); and (2) a ``fibril region'' at $x=[25,35]$\,Mm above a stronger flux concentration (containing a dynamic fibril). Thus, region (1) and (2) represent a plage periphery and a plage region respectively, as appropriate to compare with our \emph{IRIS-ALMA} observations of plage and its periphery near disk center. 
The viewing geometry chosen was for an ``observer'' looking from above down on the domain (i.e., assuming a LOS along the vertical direction in the simulation). In Figure~\ref{FIG_4} we show space-time plots (hereafter, $x-t$) for \emph{ALMA}/Band6 (panel c), \ion{Mg}{2} in a wavelength range of 0.7\,\AA\ centered at the k3 rest wavelength (2796.35\,\AA; panel e), and \ion{Si}{4} 1393\,\AA\ (panel g). We assume that the observed spicule’s orientation is not such that our LOS intersects it perpendicularly over its length, as the latter seems an extreme case for its orientation (likewise for the case where the spicule is viewed along its axis). Thus, our geometry in the model seems reasonable for the interpretation of the observations.”

\section{An Empirical Method for Determining and Characterizing Areas of Plage}\label{plagemethod}

Chromospheric plage could be defined as the region of high chromospheric intensities above magnetic spatial distributions, with stronger magnetic flux than in the quiet Sun but weaker than that of sunspots and photospheric pores. An observer can determine such areas either (a) ``by eye'' and cutting out a region manually (e.g., as in  \citealt{Carlsson_etal_2015}), or (b) in a quantitative manner, i.e., by selecting a certain intensity threshold for a chromospheric image, or even an area of moderately intense magnetic fields. However, since we are interested in quantitative comparisons between different observables in plage regions, care must be taken to exclude features that are not classified as plage. Thus we should exclude: (i) photospheric pores, which may often form sporadically in plage by random convergence of unipolar magnetic fields; and (ii) dark fibrils or other small and cool filamentary structures often seen in the vicinity of sunspot penumbras/superpenumbras. Therefore, in the present study we consider plage to be the hot magnetic canopy above photospheric magnetic concentrations typical for plage. We have been cautious to not include other elements such as pores or the superpenumbra from nearby spots. 

The flux-segmented (or thresholded) magnetograms were produced by clipping values at $\pm$0.1\,kG to distinguish between strongly and weakly magnetized areas. We use the segmented magnetograms as a visual guide to aid the determination of plage in our chromospheric observables. This confirms that the FOV is naturally split into two distinct plage regions (one north and one south in the FOV) separated by a weakly magnetized region in the middle. With this in mind and given (i) the several linear-like structures (and spicule; see Paper I) present in that quieter area, and, (ii) due to the small size of the raster along the $x$-direction (5$\arcsec$), combined with (iii) the knowledge that plage and magnetic fields reside at the west just outside the raster FOV (Figure~1b of Paper I), we conclude that this middle region is part of the \emph{periphery} of that same plage. We also note that the entire plage is at sun-center angle of $\approx20\degr$. Here, our common \emph{ALMA}/\emph{IRIS} FOV contains just the east boundaries of this plage region. This simplifies our task in determining the true chromospheric boundary of our plage on only one side. However, we caution of possible systematic geometric offsets between the boundaries of extended plage regions, e.g., if they were defined in the photosphere (via a thresholded magnetogram maps), and used to describe the boundaries of the higher-lying chromospheric plage area.


We inspected imaging in the continuum from \emph{IRIS}/SJI \ion{Mg}{2} 2832\,\AA\ which clearly shows the existence of a photospheric pore within the north part of plage in the common \emph{IRIS}-\emph{ALMA} FOV. The pore is persistent for the most part of the \emph{IRIS}-\emph{ALMA}/Band6 co-observations. Here we caution that excluding the pore from the chromospheric plage pixels with the use of a threshold on magnetogram maps is not a straightforward task. A pore's area determined at the photosphere is likely a smaller area than the pore's associated area at chromospheric heights due to lateral expansion/``fanning'' of the pore's magnetic field with height. To properly remedy this issue, we employ the \emph{ALMA}/Band6 maps\footnotemark\ segmented at a low threshold $T_b\ge6,500$\,K. This approach effectively removes the chromospheric counterpart of the pore in plage.
\footnotetext{We note that any of the Mg\,{\sc II}, Si\,{\sc IV}, or C\,{\sc II} raster maps from \emph{IRIS} could be used equally well to remove the chromospheric counterpart of the photospheric pore for a choice of threshold. However, since \emph{ALMA}/Band6 maps are produced with an irregular resolution element, i.e., the so-called ``beam'', here we make a mask from Band6 to further restrict the accidental inclusion of lower $T_b$ values in our comparisons due to spatial smearing from the beam.}
The final result from the application of our method can be seen in the top panel of Figure~\ref{FIG_CII}, where we overplot on the \emph{SDO}/HMI map the regions of plage within red contour and with orange contours showing the excluded area above the pore region. The plage periphery is the area within the blue colored contour where linear-like structures and spicules are seen to develop against a significantly darker (and quiet-sun-like) background (see Paper I).

We summarize the observational quantities and requirements to define plage:

\begin{itemize}
	
	\item photospheric magnetogram to be used as a guide (threshold choice at $\pm$0.1\,kG, although similar results can be found for even lower thresholds),
	
	\item continuum maps to properly identify areas of pores within plage or sunspot penumbras,
	
	\item chromospheric intensity maps (e.g., \ion{Mg}{2} or any other chromospheric observable) to remove pores with proper intensity thresholds above the identified regions of pores,
	
	\item avoid the vicinity of well-developed sunspots, i.e., regions of sunspot superpenumbra which typically extend further out than the sunspot penumbra. 
	
\end{itemize}

By carefully considering these criteria we can determine ``clean'' regions of chromospheric plage. As we show in the following paragraphs, when these are not taken into consideration simultaneously, sources of bias appear which lead to discrepancies between our results and previous studies.

\section{Analysis and Results}\label{results}


\subsection{Understanding the Origin of the Similarities Between \emph{ALMA}/Band6 and \emph{IRIS} Observables}\label{RESULTS_SPICULE}

The high degree of similarity between \emph{ALMA}/Band6 and \ion{Si}{4} observables is illustrated in Figure~\ref{FIG_2}. In that figure we point to locations in the FOV where \emph{ALMA}/Band6 emission appears in anti-correlation with \ion{Mg}{2} k intensity patterns (boxed areas in rasters of Figure~\ref{FIG_2}). Here we perform a quantitative analysis on the degree of similarity between \emph{ALMA}/Band6 and other chromospheric and transition region observables from \emph{IRIS}, to better understand and constrain the diagnostic potential of \emph{ALMA}/Band6 as a tool to measure the temperature of chromospheric plasmas. As we mentioned in \S~\ref{intro}, the actual formation height of \emph{ALMA} emission is not well known. In particular, to link this work with previous studies, we first carry the analysis using (a)  observables time-averaged over the entire time-series (\S\ref{STATIC}) and we discuss the discrepancies with previous studies. Then we use (b) observables without averaging in time to also consider the time-evolution (\S\ref{timedepend}). We then follow with results of our analysis from the model by  considering both the time-evolution and the height-wavelength dependence of the \emph{IRIS} observables (\S\ref{morphology_model}).

\subsubsection{Quantification of Morphological Similarities Between Observables Time-Averaged Over the Entire Image Series}\label{STATIC}

Here, we perform an intercomparison between wavelength-integrated rasters in \ion{Si}{4} 1393\,\AA, and \ion{C}{2} 1335\,\AA, \ion{Mg}{2} k 2796\,\AA\ and also with \emph{ALMA}/Band6. 

Comparison between the optically-thick observables \emph{ALMA}/Band6 $T_b$ and \ion{Mg}{2} is typically done with \ion{Mg}{2} expressed as radiative temperature, $T_{\rm rad}$, in units of temperature [K]. Previous studies (e.g., \citealt{Bastian_etal_2017, Bastian_etal_2018, Jafarzadeh_etal_2019}) found that the intensity of \ion{Mg}{2} k2 or h2 peaks correlates with mm-emission from \emph{ALMA}/Band6 observations, supporting the expectation that \emph{ALMA}/Band6 emission forms at mid-to-low chromospheric heights. To compare our study with previous works we perform double gaussian fitting \citep{Schmit_etal_2015} for the \ion{Mg}{2} raster data and produce maps representing each feature of the k and h lines, i.e., k2v, k2r, h2v, h2r, k3, h3. We compute $T_{\rm rad}$ with eq.~(\ref{planck}) for each of these maps. We also consider the wavelength-integrated \ion{Mg}{2} k quantity we produced and used in the previous sections via eq.~(\ref{planck}) and get $T_{\rm rad}$ at each wavelength position of the rasters separately (between $\Delta\lambda$=0.7\,\AA\ from line center). We then produce the average quantity of $T_{\rm rad}$ representing the wavelength-integrated \ion{Mg}{2} k data, by taking the average of $T_{\rm rad}$ produced in that range. The choice of this $\Delta\lambda$ offers the benefit of including all line features of \ion{Mg}{2} k (i.e., k2v, k2r, and k3) without extending too much into the line continuum. 

For the wavelength-averaged optically-thin observables, computing $T_{\rm rad}$ is physically meaningless. We keep the observed values expressed in arbitrary intensity units [DN s$^{-1}$]. For \ion{Si}{4}, the integration was performed for each frame in these rasters in a wavelength range of 0.2\,\AA. In order to increase the S/N in the \ion{C}{2} raster we sum both lines (each centered at 1334.5\,\AA\, and 1335.7\,\AA), and then integrate the sum in wavelength over 0.2\,\AA. The \ion{C}{2} (not always optically-thick so we keep it in [DN s$^{-1}$]) rasters suffer from low counts making the presence of hot pixels more  impactful in the statistics, with plenty of hot pixels being visible in the map from the average \ion{C}{2} image series. We determined that the hot pixels can be extracted easily, since their values exceeded the values from persistent structures owing to real \ion{C}{2} signal. Thus, at each frame, we remove any hot pixels exceeding 10 DN s$^{-1}$ and substitute the resulting missing pixel values via linear interpolation from values of the immediate neighboring pixels. While significant noise was still present in each frame due to low photon counts, the inspection of the time-averaged map before and after the removal of hot pixels showed that the S/N was improved satisfactorily. 

A common approach in multi-wavelength studies utilizing data from different observatories/instruments is to ensure that any time-differences in the image series between different observables are: (1) properly matched/synchronized; and (2) small enough and appropriate for addressing particular science questions. Both are required to effectively ``freeze-in-time'' the plasma evolution between all the different wavelengths. This becomes a serious concern in studies of the highly dynamic chromosphere, such as the one we report in the present work. Here, we match our \emph{ALMA}/Band6 2\,s-cadence series by composing ``rasters'' that match the time of each slit sampling position to within $\pm$1\,s. We have also performed the analysis presented here by selecting the frame at the time corresponding to the middle of each raster scan (raster cadence 26\,s, resulting to $\pm$13\,s time-difference) but found no significant change in our results (i.e., of order $\sim$ 1\%). This presents an improvement as compared to previous studies (e.g., \citealt{Bastian_etal_2018, Jafarzadeh_etal_2019}), where the minimization of time-differences was limited due to the data series used, resulting in a highly variable time-matching between \emph{ALMA-IRIS} observables (i.e., 0.5-2\,min). In particular, \citet{Jafarzadeh_etal_2019} acknowledge that significant evolution may be ongoing during this period of time between the sampling of the \emph{ALMA-IRIS} observables, and also reported findings by a separate analysis where the time difference was strictly chosen to be 0.5\,min (marginally improving the agreement between \emph{ALMA-IRIS} observables). Here, with a maximal difference of $\pm$1\,s we ``freeze'' the plasma evolution consistently and successfully between each observable over our entire \emph{ALMA-IRIS} image series. However, since the \ion{C}{2} rasters have low photon counts, in this subsection we restrict the comparison between time-averaged rasters over the entire \emph{IRIS}-\emph{ALMA}/Band6 common time series (for a time-dependent study \emph{ALMA}/Band6 and \emph{IRIS} \ion{Mg}{2} k and \ion{Si}{4} see next subsection \S~\ref{timedepend}). 

To perform a fair comparison with \emph{ALMA}/Band6, we degrade the wavelength-integrated \emph{IRIS} raster maps by convolving them with the Band6 beam size and respective position angle at each time frame of our series. Finally, we apply 4-pixel binning along the slit direction for each observable to additionally increase the S/N. The resulting Band6-beam-degraded and time-averaged \emph{IRIS} maps (before and after 4-pixel binning) are shown in the top panels of Figure~\ref{FIG_CII}. In the panels before binning (top left) we can see some cosmetic artifacts, namely a dark line due to the fiducial point which blocks the light in the slit, in addition to a linear-shaped intensity depression (mostly seen in the FUV observables), which appears similar to the shadow produced by the fiducial point. In our analysis, we mask out and exclude these two rectangular areas (see dark bands along the $x$-direction in the 4-pixel binned maps). 

\paragraph{Quantitative Comparison Between Time-Averaged Observables in Plage and its Periphery}\label{comparison}
\

At first sight, the similarity between \ion{C}{2} and \emph{ALMA}/Band6 is striking. We proceed by calculating the linear correlation coefficent (Pearson r, hereafter ``C.C.'') between each of the four observables (i) inside the plage region, (ii) inside the region containing the periphery of plage, and (iii) for the full FOV in the rasters mixing together plage with its periphery, first with the presence of the pore (the respective C.C. values for each region are shown within black boxes at the top of each scatter plot). Then we also calculate the C.C. for regions (i) and (iii), after the pore area is excluded in each of the observables (values shown within the dashed orange boxes in each scatter plot; region (ii) is obviously unaffected from the removal of the pore pixels). The inclusion of this low temperature \emph{ALMA}/Band6 region in the calculation of the C.C. for plage and for the full FOV can be readily seen to influence the values presented in the scatter plots in Figure~\ref{FIG_CII}abc. The pixels in the region above the pore produce a clear ``spire''-like feature or a ``tail of points'' towards low \emph{ALMA}/Band6 $T_b$ values (i.e., $4,000\le T_b\le6,500$\,K, i.e., our $T_b$ threshold choice for producing the pore mask is fully consistent with representing this feauture). Removing those pixels with the application of the pore mask improves the C.C. for the full FOV and plage significantly (red points marked with an orange ``$\times$'' symbol). However, we caution that the C.C. obtained over the full FOV leads to different results since it mixes plage regions with much quieter regions in the plage periphery, i.e., regions with very different physical conditions. 

We first focus our analysis and discussion for the plage (excluding the pixels above the photospheric pore) and periphery regions separately and keep the analysis for the full FOV to link our work to previous studies. To facilitate the presentation of the results, in the bottom panels of Figure~\ref{FIG_CII} we organize the various C.C. values obtained from each combination of the observables with the aid of correlation matrices. The correlation matrix for plage shows that \ion{C}{2} is highly correlated with all other observables (C.C. ranging 0.73 to 0.79). In contrast to that, the observable with the lowest correlation between observables in plage is \ion{Mg}{2} k $T_{\rm rad}$. \emph{ALMA}/Band6 and \ion{Si}{4} and \ion{C}{2} form a triad with the highest C.C.. Perhaps this should not be a surprise, since these high values are fully consistent with \ion{Si}{4} and \emph{ALMA}/Band6 forming close to each other, thus probing similar structures. In addition, we emphasize that the \ion{Mg}{2} $T_{\rm rad}$ is averaged over 0.7\,\AA\ so that it mixes information from a wide range of heights. We note that if we use the wavelength-averaged \ion{Mg}{2} k time-averaged map expressed in intensity [DN sec$^{-1}$] instead of $T_{\rm rad}$ in [K] our results in the correlation study we present here do not change beyond a few percent. In addition we emphasize that if we impose thresholded magnetogram masks for the plage areas the results do not change beyond 3-5\% which further confirms the rubustness of our methodology described in \S~\ref{plagemethod}.

The similarity between \ion{C}{2} with \ion {Si}{4} has been mentioned previously in visual comparisons between \emph{IRIS} \ion{C}{2}, \ion{Si}{4} and \ion{Mg}{2} rasters (e.g., \citealt{Rathore_etal_2015a}). In a study by \citet{Rathore_etal_2015b} (e.g., Figure 17 therein), \ion{Si}{4} was found to form consistently higher (having normally a formation height in the transition region, around $T\approx$80,000\,K) than \ion{Mg}{2} and \ion{C}{2}, although \ion{C}{2} was found to form at heights either above or below \ion{Mg}{2} k3. However, it was also noted that \ion{C}{2} primarily formed above the formation height of \ion{Mg}{2}. 

Our results in the correlation matrix for plage (Figure~\ref{FIG_CII}) support this finding, given that \ion{C}{2} shows very high correlation with \emph{ALMA}/Band6 and \ion{Si}{4}, the later being understood as all these observables have formation heights relatively close to each other, effectively sampling the conditions along similar parts of structures above plage. \ion{Mg}{2} k $T_{\rm rad}$ seems poorly correlated with all other observables but \ion{C}{2}, suggesting that \ion{C}{2} forms above \ion{Mg}{2} k but between \ion{Mg}{2} k and \emph{ALMA}/Band6 and \ion{Si}{4}. 
To this we should add that the interesting anti-correlation seen between maps of \ion{Mg}{2} k and \emph{ALMA}/Band6 and \ion{Si}{4} in certain locations of the FOV (Figure~\ref{FIG_2}; boxed regions in plage), presumably due to enhanced absorption, may have the effect of weakening the correlation of \ion{Mg}{2} k with the other observables. 

The correlation matrix for the periphery of plage shows a similar picture. The C.C. between \ion{C}{2} and the other three observables remains the highest as compared to any other combination of three (out of four) observables. The significant strength of the correlations with \ion{C}{2} for the periphery appears consistent with the results of \citet{Rathore_etal_2015b}, which place the formation height of \ion{C}{2} higher than \ion{Mg}{2} and thus closer to \ion{Si}{4} in the fibril regions (Figure 18 therein). Note, however, that \citet{Rathore_etal_2015b} did not average \ion{Mg}{2} k over 0.7\,\AA. The periphery of plage here contains several linear-like structures, which suggest similar geometry as in the simulations shown in \citet{Rathore_etal_2015b} (e.g., see in our Figure~\ref{FIG_CII} the persistent thread-like structures in the time-averaged maps). On the other hand, this region also exhibits low signal in \ion{Si}{4} and \emph{ALMA}/Band6. This may explain the significantly lower correlation between \ion{Mg}{2}, \ion{Si}{4}, and \emph{ALMA}/Band6 as compared to that in plage, since \ion{Mg}{2} intensity seems more diffuse in that region. For completeness, we also note here that a study of correlations between \ion{C}{2} $T_{\rm rad}$ and \emph{ALMA}/Band6 $T_b$ was performed by \citet{Jafarzadeh_etal_2019} and also found a high Pearson C.C. of 0.83, although it was mentioned there that the origin of this high correlation with \emph{ALMA}/Band6 was not understood. 

Since our results are in contrast with previous studies we further discuss the reasons behind the discrepancies in a separate paragraph \S~\ref{discrepancies}.

\subsubsection{Time-Dependent Quantitative Study of Morphological Similarities Between Observables}\label{timedepend}

We note that during the time range of co-observations with \emph{ALMA}/Band6, significant evolution was ongoing over the entire FOV. To get a clearer picture and to assess the similarities by considering the dynamic evolution in the common FOV, we further focus this analysis on \ion{Si}{4} and \ion{Mg}{2}, which, thanks to the higher photon counts, can be used to explore correlations with \emph{ALMA}/Band6 at each time-step of our comprehensive dataset. 

In order to determine which regions in the common \emph{IRIS}-\emph{ALMA} FOV exhibit high morphological similarities between our set of observables, we split the FOV in sub-regions, or ``correlation bins'', and we measure the C.C. between the different observables at each time step of our series. In the top panels of Figure~\ref{FIG_RASTER_CORREL_STACK} we show the size of the correlation bins on time-averaged, wavelength-integrated maps degraded with the \emph{ALMA}/Band6 beam and position angle. Furthermore, we distinguish the sign of the C.C. -- positive C.C., denoting positive correlation of the intensities, and negative C.C., representing anti-correlation. The latter will allow us to locate when and where such anti-correlation between (1) \ion{Mg}{2} k, and (2) either \ion{Si}{4} or \emph{ALMA}/Band6 occurs. The time-evolution of the C.C. per bin is presented in the form of time-plots in the bottom panels in Figure~\ref{FIG_RASTER_CORREL_STACK}. For reference, we also show time-plots of the average time-evolution within the correlation bins for each of the observables as average intensity per bin. Note that for the calculation of the C.C. per bin per time-step we did not perform spatial averaging and we used all the individual pixels within each correlation bin at each time step. Also note that we have degraded the \emph{IRIS} observables with the \emph{ALMA}/Band6 beam size and position angle. 

The resulting time-plots in Figure~\ref{FIG_RASTER_CORREL_STACK} vividly highlight the similarities and the differences between the observables. Again, as before, the raster series is integrated along wavelength centered on each line's rest wavelength. Remarkably, the FOV of \ion{Mg}{2} k and \ion{Si}{4} is split in locations of positive correlation and anti-correlation, confirming our initial visual determination of some locations of anti-correlation between these two observables (e.g., Figure~\ref{FIG_2}; boxed regions in rasters). In addition, at times, there are certain locations where there is strong correlation between the intensities of \ion{Mg}{2} k and \ion{Si}{4}, particularly in the region containing the periphery of plage with the linear-like structures and the Type-II spicule. However, the most remarkable finding is the sporadic correlation (both in terms of intensity and time persistence) of \emph{ALMA}/Band6 with \ion{Mg}{2} k and the very high and more persistent correlation between \emph{ALMA}/Band6 with \ion{Si}{4} across the entire FOV, with only a few instances of anti-correlation. \ion{Mg}{2} k is found in anti-correlation with \emph{ALMA}/Band6 in several locations in the FOV. The strong correlation between \emph{ALMA}/Band6 and \ion{Si}{4} suggests that the spatial extent of bright features -- as projected on the plane of the sky -- is similar between these two observables, supporting that the geometric height of  line formation of \emph{ALMA}/Band6 and \ion{Si}{4} is similar. 

Conversely, \ion{Mg}{2} k and \emph{ALMA}/Band6, even though they are nominally expected to have similar plasma temperature sensitivity, appear to sample different geometric heights in the solar chromosphere. This finding is consistent with the synthetic data from the model (Figure~\ref{FIG_4}). Previous works (e.g., \citealt{Bastian_etal_2017}) considered that \emph{ALMA}/Band6 forms at the mid-range of \ion{Mg}{2} formation heights, but as we show later (\S~\ref{morphology_model}) this does not appear to be the case for our dataset. This is also supported by our study on time-averaged observables in the previous subsection \S~\ref{comparison}. Note, however, that we use wavelength-averaged rasters, and by averaging in wavelength the diagnostic information regarding the formation height of the \ion{Mg}{2} k line is biased to lower heights as only \ion{Mg}{2} k3 forms at the top of the chromosphere. In fact, in the \ion{Mg}{2} k $x-t$ plot for $\tau=1$ of Figure~\ref{FIG_4}f for the maximum geometric height, we find a good match with geometric heights for \emph{ALMA}/Band6 $\tau=1$. However, \ion{Mg}{2} k3 forms due to absorption -- and thus, rasters in k3 are capturing the maximum absorption in that line. It is therefore our expectation that when \ion{Si}{4} and \emph{ALMA}/Band6 show emission in dynamic plage structures (or spicules), \ion{Mg}{2} k3 (which would be closer to \emph{ALMA}/Band6 heights) has low intensity due to enhanced absorption. This can lead to anti-correlation with the other observables. We address this in the next section \S~\ref{morphology_model}. 

For reasons of completeness, we perform this analysis on the synthetic observables produced from our model. In this case, since the simulation is 2.5D, the synthetic observables can be likened to a ``sit-and-stare'' \emph{IRIS} observation, capturing the evolution across a ``static 1D slit''. In Figure~\ref{FIG_SIM_CORREL_STACK} we show the results. In the top panels we show how we split the domain in correlation bins (here, the bins are essentially 1D, arranged along the simulation domain's $x$-direction at each time-step). As in the previous sections, the data have been degraded from the simulation's scale size, 14\,km (grid point)$^{-1}$, via gaussian convolution to adopt the nominal spatial resolution of \emph{IRIS} rasters (0$\farcs$16 pix$^{-1}$ along y-direction) and \emph{ALMA} (degraded with the Band6 maximum beam size of $0\farcs8$). Then we further degrade the synthetic \emph{IRIS} data by convolving the \emph{ALMA}/Band6 beam size. In addition, we have masked out the location of emerging flux, which effectively separates the ``spicules region'' at the top of the FOV from the ``plage region'' at the bottom. The photospheric $B_z$ shows significantly higher magnetic field strength for the plage region as compared to that in other locations of the domain (i.e., of order $\approx$100\,G). Since we used the same ``correlation bin'' width with the observations in Figure~\ref{FIG_RASTER_CORREL_STACK}, only a few bins cover the domain in these time-plots. However, qualitatively, we get the same picture as before. 

In Figure~\ref{FIG_SIM_CORREL_STACK} we highlight the region of spicules with a solid ellipse and we use a dotted ellipse for the plage. For \ion{Mg}{2} k and \ion{Si}{4} we see primarily anti-correlation for the plage region (compare PC1 dashed ellipse with solid PA1) and an alternation of correlation and anti-correlation for the spicule region (compare solid ellipses SC1 and SA1). The latter seems consistent with  Figure~\ref{FIG_RASTER_CORREL_STACK} where the plage region in  \ion{Mg}{2} k and \ion{Si}{4} was more anti-correlated and showed a more intermixed correlation/anti-correlation for the plage periphery/spicule region (i.e., the anti-correlation plot in the observations shows less strong anti-correlation as compared to the plage regions north and south of the raster's FOV). For \emph{ALMA}/Band6 and \ion{Si}{4} we primarily see strong correlation for the plage (Figure~\ref{FIG_SIM_CORREL_STACK}; compare dotted ellipse PC2 with PA2), and some alternation between correlation and anti-correlation (with clear correlation during the time of the network jet, after $t$=3,600\,s) for the spicule region (compare area of solid ellipses SC2 with SA2). Last, for \ion{Mg}{2} k and \emph{ALMA}/Band6, we see anti-correlation for the plage region (compare PA3 with PC3), but a somewhat sporadic occurrence of correlation for the plage periphery/spicule region (compare area in solid ellipses SC3 with SA3; also positive correlation during the time of the network jet) intermixed in areas of anti-correlation, again in general agreement with previous comparison between \emph{ALMA}/Band6 and \emph{IRIS} obervations (Figure~\ref{FIG_RASTER_CORREL_STACK}).

\subsubsection{Explaining the Morphological Similarities in the Synthetic Observables}\label{morphology_model}

In order to understand this interesting correlation in the observations we explore the time evolution in the emission and formation height of the synthetic observables for the region with spicules and plage (Figures~\ref{FIG_SIM_EMISSIV_SPICULES} and~\ref{FIG_SIM_EMISSIV_PLAGE}, respectively; also see Paper I for a detailed analysis on the Type-II spicule). For our discussion here, we select three representative time-steps along the evolution of a region with spicules and plage in the simulation. In addition, we select three wavelength positions for \ion{Si}{4} and \ion{Mg}{2} k (taken in velocity space at -13, 0 and +13 km s$^{-1}$ from the line cores). By doing so, we capture the wavelength dependence of the formation height of \ion{Si}{4} and \ion{Mg}{2} k during the ascending and the descending phase of the spicule's evolution (\bf see also Figure 8 in Paper I)\rm. At each time shown in Figures~\ref{FIG_SIM_EMISSIV_SPICULES} and~\ref{FIG_SIM_EMISSIV_PLAGE} we provide a wavelength-space plot (hereafter, ``$\lambda-x$'') for \ion{Mg}{2} k (i.e., full spectral profile along the different positions in the simulation domain), which add relevant information in support of our interpretation presented in this subsection. 

At the ascending phase of
spicule 1 ($t$=3,350\,s in Figure~\ref{FIG_SIM_EMISSIV_SPICULES}, position $x=[42,44]$\,Mm; spicule 2 has not started forming yet) we get emissivity in \ion{Si}{4} delineating the spicular column at the blue wing (-13 km s$^{-1}$; panel a; dark green-to-white color outlining spicule 1), which closely corresponds to the height of \emph{ALMA}/Band6 emission (shown with dark-red-to-white color in all panels). At the same time, in the area where spicule 2 would eventually develop (position $x=[40,42]$\,Mm) we can see significant \ion{Mg}{2} k intensity in the blue wing (plotted at $\tau=1$) (Figure~\ref{FIG_SIM_EMISSIV_SPICULES}a). In Figure~\ref{FIG_SIM_EMISSIV_SPICULES}d we show the $\lambda-x$ plot for this time with the \ion{Mg}{2} k spectrum at each $x$-position of the domain, with clear RBE-signatures in the location of formation of spicule 2. Also, in the blue wing of \ion{Si}{4} (Figure~\ref{FIG_SIM_EMISSIV_SPICULES}a) we see a front at the tip of the structure delineated by \ion{Mg}{2} k intensity (heights around z=2.7\,Mm). This is consistent with the effects of a traveling shock waves in the chromosphere before the full development of a spicule. 

At the intermediate time-step shown in Figure~\ref{FIG_SIM_EMISSIV_SPICULES} ($t$=3,580\,s; middle column), spicule 2 has been fully developed and has almost reached its maximum elongation (z=5\,Mm from the photosphere); most of \ion{Si}{4} emissivity now comes at the rest wavelength of the line (panel f; dark green-to-white color outlining spicule 2). Similarly, at the time of maximum elongation, \emph{ALMA}/Band6 emission delineates the body of spicule 2 (note the close matching of geometric heights of line formation between \emph{ALMA}/Band6 and \ion{Si}{4}; dashed ovals along spicule 2). 
The spectral profile of the optically thick \ion{Mg}{2} (Figure~\ref{FIG_SIM_EMISSIV_SPICULES}h) shows that the k3 (``dark lanes'' in the $\lambda-x$ plot highlighted with white dotted lines) has shifted at different wavelength positions accross the spicule (e.g., panels f and g), and the respective geometric height of $\tau=1$ at those different wavelength positions delineates different parts of the spicule (see dashed ovals in panels f and g). In fact, the spicule in \ion{Mg}{2} appears as a dark feature (as compared to other locations in the domain), since the line is in absorption. This is either due to the increased opacity or due to the lower source function.

In the last time-step shown here ($t$=3,620\,s), spicule 2 has already begun to recede and the occurrence of the network jet brightening along the spicule strongly enhances the emission in \ion{Si}{4} at the core and in the red wing (panels j and k). The impulsive heating of the plasma in the spicule is so intense that forces \emph{ALMA}/Band6 emission to come from lower heights where the plasma is cool enough (the spicule's height seen in \emph{ALMA}/Band6 effectively drops from z$\approx$5\,Mm to z$\approx$3\,Mm; shown with arrows in panels j). At the same time, while significant intensity in \ion{Mg}{2} k emanates from low geometric heights (dashed ovals between z=1 - 1.5\,Mm; panels i,j) at the blue wing and at line core, the \ion{Mg}{2} k $\tau=1$ height in the red wing (panel k) is much greater. There, we can see that \ion{Mg}{2} k $\tau=1$ roughly traces the length of the spicule, albeit in low intensity as compared to other locations with lower geometric heights (note that the same intensity range is used for each of the panels). This, again, is a manifestation of increased absorption. 
Here, Figure~\ref{FIG_SIM_EMISSIV_SPICULES} (panel l) reproduces this behavior during the time of the network jet (pointed by an arrow in that panel). 

The simulation captures clearly that the high correlation between spatially-resolved structures seen in \emph{ALMA}/Band6 and \ion{Si}{4} is primarily due to the fact that the respective emissions emerge from similar parts of the same structure, both largely delineating the spicular column. Therefore, when observed at the plane of the sky (looking from the top of the simulation domain in this case, or, in other similar LOS, e.g., off the vertical) the spicule would manifest in both said observables and the high-intensity features would show up largely as spatially correlated. We cannot say the same, however, for \ion{Mg}{2} k, as the $\tau=1$ geometric height varies at different wavelength positions. However, we note a characteristic trend: during the ascending (descending) phase of the spicule, the $\tau=1$ height at the blue (red) wing roughly delineates the spicule, albeit as a dark structure. There is also the following possibility: if the viewing angle (LOS) was tilted, say by 30$\degr$ off the vertical towards small $x$ (left side of the domain shown in Figure~\ref{FIG_SIM_EMISSIV_SPICULES}j,k), we would be seeing this effect to develop as: (i) bright \ion{Si}{4}; (ii) bright \emph{ALMA}/Band6; and (iii) dark \ion{Mg}{2} k, since the high intensity in \ion{Mg}{2} k would be at the root of the spicule and the emission of (i) and (ii) would be projected against an area of low \ion{Mg}{2} k background emission (e.g., for $x>$44\,Mm). This example illustrates one of the possible reasons that at certain locations in the observations \ion{Mg}{2} intensity is anti-correlated to both \ion{Si}{4} and \emph{ALMA}/Band6.

The relationship between the ascending/descending phase of mass motions with Doppler shift and emissivity in \ion{Si}{4} is also seen in the plage region of the simulation (Figure~\ref{FIG_SIM_EMISSIV_PLAGE}). At $t$=3,400\,s a dynamic fibril shoots mass upwards ($x=[27,29]$\,Mm), where in the blue wing a ``front'' of \ion{Si}{4} emissivity is closely followed by emission in \emph{ALMA}/Band6 (panel a; dashed oval). \ion{Mg}{2} k intensity comes from greater geometric heights but in absorption, again, in contrast to other locations in the domain, which are brighter but at much lower geometric heights ($z\approx$0.5-1\,Mm). Similarly, relatively low intensity is seen at the nominal rest wavelength of \ion{Mg}{2} k (panel b). This, again, pinpoints the reasons behind the anti-correlation we noted between \ion{Mg}{2} k and the well-correlated pair of \emph{ALMA}/Band6 and \ion{Si}{4}, both in the simulation and the observations of plage. Moving forward at the times of the other two time-steps ($t$=3,510\,s and 3,580\,s), we see a consistent evolutionary pattern between \emph{ALMA}/Band6 and \ion{Si}{4}. That is, when the bulk of the mass in the dynamic fibril stalls, we get emission from \ion{Si}{4} line-core (panel f) and when the mass is receding back to the surface, \ion{Si}{4} emits in the red wing, again, closely followed by \emph{ALMA}/Band6 emission (panel k). We also note that at those times high intensity in \ion{Mg}{2} k comes primarily from lower heights in the atmosphere (see panels i, j, and dashed oval in panel k).

Therefore, we conclude that \emph{ALMA} is sensitive to the cool-to-warm plasma existing at the highest parts of either spicules and dynamic fibrils but just below their tips. These locations are subject to shocks or other cooling/heating mechanisms (such as ambipolar heating or cooling by adiabatic expansion), which may raise the plasma to high temperatures, eventually causing it to emit in \ion{Si}{4} ($T\approx$80,000\,K) or even lower temperatures. Subsequently, the temperature drops down to a level that \emph{ALMA}/Band6 is sensitive to (8,000-10,000\,K), in geometric heights not far from those of transition region temperatures. This finding gives insights on the multi-thermal nature of spicules \citep{Chintzoglou_etal_2018}. All the above result in the high similarity between \ion{Si}{4} and \emph{ALMA}/Band6 seen in the observations. With regards to \ion{Mg}{2} k, depending on the viewing angle and on how clearly such effects are seen against the background, comparisons with the other observables may show a loss of correlation, or, if there is regularity in the appearance and positioning of such structures within the FOV, anti-correlation may also arise (i.e., bright features in one observable/pass band ``complementing'' dark structures in another). This seems to be reasonable when spicules or dynamic fibrils are bright in \ion{Si}{4} and \emph{ALMA}/Band6 but manifest as low intensity features in \ion{Mg}{2} (due to enhanced absorption and/or due to low intensity in comparison to \ion{Mg}{2} k intensities from other locations; for the case of dynamic fibrils this may be related to similar effects that were seen in space-time plots of bright grains in the work of \citealt{Skogsrud_etal_2016}). 

\subsection{Measuring the Temperature Increase in the Aftermath of Shocks Above Plage}\label{SHOCKS}

Apart from the weak magnetic flux plage periphery, the common \emph{IRIS} and \emph{ALMA}/Band6 FOV contains parts of a strong magnetic flux plage region that is very dynamic. In this subsection we focus on the dynamics seen in the plage region and we explore the potential of \emph{ALMA}/Band6 observations in measuring the plasma temperature and its time evolution in regions dominated by the passage of chromospheric shocks.

Here, with \emph{IRIS} observing in \ion{Mg}{2} k and \ion{Si}{4} we can see the chromospheric shocks as they propagate higher in the chromosphere/transition region. Focusing on the north part of the common \emph{IRIS} and \emph{ALMA}/Band6 FOV we see a lot of recurrent activity as brightenings above the plage, and also some plane-of-the-sky motions of bright dynamic fibrils. Chromospheric plage exhibits features known as dynamic fibrils, driven by slow-mode magnetoacoustic shocks which pervade the plage region \citep{Hansteen_etal_2006, DePontieu_etal_2007a, Langangen_etal_2008}. From our 34\,min-long time series we calculate the autocorrelation at each pixel in the \ion{Mg}{2} FOV to determine locations of activity segmented by the characteristic lifetimes of the signal, such as intensity modulations caused by chromospheric shocks. \citet{DaSilvaSantos_etal_2019} have reported a periodicity for shocks in \ion{Mg}{2} k of 3.5 to 4\,min in plage. In the areas above plage at \ion{Mg}{2} k2v we found autocorrelation times of $\approx$150-200\,s but also some even ``slower'' locations of $\approx$300\,s. Using this autocorrelation map as a guide we select a $1\arcsec\times1\arcsec$ sub-region (corresponding to a 3$\times$6 pixels for \emph{IRIS} rasters and 6$\times$6 pixels for \emph{ALMA}/Band6) that is well within plage (position $(x,y)=(4,17)\arcsec$ in the raster FOV of Figure~\ref{FIG_2}) in order to explore if \emph{ALMA} can be used to study chromospheric shocks. Within the selected region, we produce $\lambda-t$ plots for \ion{Si}{2} and \ion{Mg}{2} k \emph{IRIS} rasters and extract the $T_b$ from \emph{ALMA}/Band6. Furthermore, to improve the contrast in the $\lambda-t$ plots we filter them with an unsharp-mask image processing operation with a 5-pixel radius. Also, to enhance weak features that were still not visible in the wings of the lines, we produced the time-derivative of the direct $\lambda-t$ maps. We present the results in Figure~\ref{FIG_8}.

As a shock passes through the chromosphere above plage, a typical behavior is seen in $\lambda-t$ plots of chromospheric lines: a blue-shifted excursion slowly drifts toward the red wing of the line, until a new blue-shifted excursion appears again, and so on, producing a ``sawtooth'' pattern in the $\lambda-t$ plot. This ``sawtooth'' pattern is seen in \ion{Mg}{2} k $\lambda-t$ with sudden increases of the intensity in the far blue wing that sweep through the dark k3 core and then reach the red wing typically with a new enhacement in the blue wing (Figure~\ref{FIG_8}b; see enhancements above yellow line in the blue wing). Similarly, blue-shifted enhancements appear in \ion{Si}{4} \citep{Skogsrud_etal_2016} in tandem with the excursions in \ion{Mg}{2} k (Figure~\ref{FIG_8}bd). We overplot the \emph{ALMA}/Band6 $T_b$ over the $\lambda-t$ plots for the same selected region above plage. Despite the data gaps in \emph{ALMA}/Band6 observations, the behavior of $T_b$ jumps is strikingly similar to the wavelength-drift trends due to the passage of the shocks in the chromosphere. In Figure~\ref{FIG_8}ac we enhance the signatures of the onsets of shocks in the time-derivative plots and mark them with arrows. In fact, when looking at the blue wing, the similarity of the $\lambda-t$ time evolution with \emph{ALMA}/Band6 is more obvious for \ion{Si}{4} than for \ion{Mg}{2} k (see prominent blueshifts in \ion{Si}{4} $\lambda-t$ plot in areas pointed by arrows 2,3,4,5 in panel d), which may have to do with the similarity of \emph{ALMA}/Band6 mm-emission formation height with \ion{Si}{4} we determined in the present paper. This result may also be consistent with \citet{DaSilvaSantos_etal_2019} who through inversions determined that during the passage of shocks the \emph{ALMA}/Band6 emission appears to emerge from lower optical depths.


Our interpretation is that the \emph{ALMA}/Band6 observations in plage are sensitive to the localized heating of the upper chromosphere/lower transition region, produced by the passage of shocks. The jumps in T$_b$ are of order 10\%-20\% increase from a baseline value of $\approx$ 7,500\,K (maximum jump at 8,500\,K). The observed decay time down to the baseline value is of order $\approx$60-120\,s. However, note that for \ion{Si}{4} $\lambda-t$ the signal in the selected pixel position in the raster occasionally becomes poor due to low photon counts in the FUV range of the \emph{IRIS} spectrograph. The 3$\times$6 pixels for \emph{IRIS} seems to improve the signal. We also note that the bright signatures of shocks seen in the rasters are not confined to one pixel location and appear to move in the plane of the sky until they fade. Again, the 3$\times$6 pixel window compensates for the most part of the plane-of-the-sky motions seen in the movie. There are times when the recurrence of brightenings took place every $\approx$120\,s (at least in the early phase of the plots; see arrows 1, 2 and 3 in panel b). This may suggest that a given location above plage may be pervaded by shocks coming from different directions which arrive at comparable timings. Such occasions may cause the periodicity patterns to modulate with shorter or even irregular periods. In addition, note that the spatial resolution of \emph{ALMA}/Band6 maps is significantly worse as compared to that of the \emph{IRIS} rasters (by approximately one order of magnitude), thus the filling factor of the shock (at the \emph{ALMA}/Band6 beam size) is less than 1 (as compared to the spatial scales resolved with \emph{IRIS}). Therefore, even though such 1000\,K-jumps in mm-emission appear consistent with heating due to shocks suggested by \citet{Wedemeyer_etal_2007}, we conclude that the true temperature enhancements may be even higher locally. Nevertheless, our work presents indications for the localized heating of the chromospheric plasma in plage regions by shocks that travel through the geometric height of formation of \emph{ALMA}/Band6 free-free emission. 

\section{Discussion on the Discrepancies with Previous Studies}\label{discrepancies}

The detailed comparison between $T_{\rm rad}$ (EUV) and $T_b$ (mm-wavelengths) in spatially-resolved morphological features has recently became possible thanks to the high-resolution and high-cadence \emph{ALMA} observations. High positive correlations for \ion{Mg}{2} $T_{\rm rad}$ and \emph{ALMA}/Band6 $T_b$ have been reported in the literature for regions of plage  (e.g., \citealt{Bastian_etal_2017, Bastian_etal_2018, Jafarzadeh_etal_2019}). However, the C.C. we found for \ion{Mg}{2} and \emph{ALMA}/Band6 (see \S~\ref{comparison}) lies in the low end of those reported in previous studies. Below we explore the methodology used in previous studies in an attempt to replicate the higher C.C. reported for \ion{Mg}{2} $T_{\rm rad}$ and \emph{ALMA}/Band6 $T_b$ and to pinpoint the cause of the discrepancies with our results.

\citet{Bastian_etal_2018} compared $T_{\rm rad}$ of \ion{Mg}{2} h to \emph{ALMA}/Band6 $T_b$ and reported a very high, C.C. = 0.80, for plage. To perform a comparison we should first discuss the differences between our work and that of \citet{Bastian_etal_2018}. First, the $T_{\rm rad}$ in \citet{Bastian_etal_2018} was produced from maps of the average $T_{\rm rad}$ of \ion{Mg}{2} h2v and h2r by also including single-peak \ion{Mg}{2} h profiles in those maps. Here, we integrated \ion{Mg}{2} k over 0.7\,\AA. The observing program that obtained the observations in \citet{Bastian_etal_2018} did not allow for a good synchronization between the \emph{IRIS} and \emph{ALMA} observables. In the present study, we obtained \emph{IRIS}/\emph{ALMA} observations with a less restrictive observing program to achieve a minimal time-difference among all observables. We also note that \citet{Bastian_etal_2018} determined plage regions with different criteria than ours (\S~\ref{plagemethod}): it was based on the visual identification of prominent morphological features in the FOV (i.e., contours roughly containing high \ion{Mg}{2} h2v and h2r average $T_{\rm rad}$ and $T_b$ for plage; \citealt{Bastian_etal_2017}; Figure 1f therein). Here we employ strict criteria for the definition of plage and its periphery, based on the methodology presented in \S~\ref{plagemethod}. 
However, given the large discrepancy in the C.C. (i.e., the C.C. reported in \citet{Bastian_etal_2018} is 40\% larger than our value), here we also calculate the $T_{\rm rad}$ in the same way as in \citet{Bastian_etal_2017, Bastian_etal_2018}. That is, we calculated $T_{\rm rad}$ for the average \ion{Mg}{2} h2v and h2r line. After doing so, we found that the C.C. in that case was even lower, i.e., C.C. = 0.52, yielding an even larger discrepancy between our results and the results reported in \citet{Bastian_etal_2018} (i.e., 54\% larger than our value). 

One possibility behind the discrepancy mentioned above is that the visual criteria employed for the definition of plage in \citet{Bastian_etal_2018} allowed the inclusion of low $T_{\rm rad}$ and $T_b$ areas in plage. For instance, the lower plage mask in Figure 1f of \citet{Bastian_etal_2017} clearly shows lower intensities for the majority of the pixels therein. Thus, such pixels should not be classified as plage, given the consideration we make in this paper (see \S~\ref{plagemethod}). The high C.C. of 0.80 that this study reported between \ion{Mg}{2} h to \emph{ALMA}/Band6 $T_b$ was obtained for the entire common FOV. We emphasize that such high correlation is not surprising if we consider the overall similarity over a diverse set of features present within that FOV, such as plage, sunspot umbra, and penumbra/superpenumbra, which ``as a whole'' appear morphologically similar between \ion{Mg}{2} h $T_{\rm rad}$ and \emph{ALMA}/Band6 $T_b$. For example, if we take the region with sunspot umbra, i.e., a morphological feature that is darker than the average intensity in the entire FOV in \ion{Mg}{2} h, it also appears dark in \emph{ALMA}/Band6 (compare Figure 1a and 1c therein). This equivalence also holds true for the plage regions, which stand out as regions of higher intensity in either chromospheric observable. Thus, mixing regions with low $T_{\rm rad}$ and $T_b$ (e.g., sunspot umbras, plage periphery, pores) together with brighter regions and treating this mix as ``plage'' may effectively increase the C.C.. A quick comparison between our correlation matrix for the full FOV (mixed) with that for plage in Figure~\ref{FIG_CII} supports our view. The mixed correlation matrix treats all plage and plage periphery points as one population (still excluding the pore). While the C.C. for \emph{ALMA}/Band6 $T_b$ with \ion{Mg}{2} k $T_{\rm rad}$ is 0.56 for plage and 0.57 for the periphery, the calculation for the full FOV increases the value dramatically, to a C.C. of 0.74, which amounts to 90\% of the value reported in \citet{Bastian_etal_2018}. The total number of pixels in the plage mask is 650 which finally becomes 578 pixels after excluding the pore region (recall that we performed binning of 4 pixels along the y-direction). The total number of points in the periphery is 480. Thus, the inclusion of an additional 80\% of plage periphery pixels (with about 2/3 showing low intensities) in the designated plage area made a clear difference in the C.C. between \emph{ALMA}/Band6 $T_b$ and \ion{Mg}{2} k. While we cannot exclude the possibility of additional factors behind this discrepancy (e.g., variability between different solar regions, different calibration methods employed in \citet{Bastian_etal_2018}, and that the pixels in IRIS and ALMA observations were not selected based on minimal time-difference constraints), including quieter pixels within plage regions may play a significant role in increasing the C.C. from a marginal value (0.56) to a value suggesting high positive correlation (0.80). 

The study by \citet{Jafarzadeh_etal_2019} focused on \emph{ALMA}/Band6 and \emph{IRIS} observations of the same region analyzed in \citet{Bastian_etal_2018}. This work also investigated relationships between $T_b$ from \emph{ALMA}/Band6 and $T_{\rm rad}$ from \emph{IRIS} \ion{Mg}{2} k and h (for each line components), \ion{C}{2}, and intensities of the optically thin \ion{Si}{4} and \ion{O}{1}. Their region of plage was defined with a quantitative method, i.e., as a chromospheric region above photospheric magnetic fields $\geq\pm$0.2\,kG, a threshold value not too different from the one we use here. Indeed our magnetic map does not change if we use either $\pm$0.1\,kG or $\pm$0.2\,kG for thresholds. However, we point out that their methodology lead to the inclusion of areas in the sunspot penumbra/superpenumbra that in \citet{Bastian_etal_2018} were excluded from their plage areas. Nevertheless, the region designated as plage in \citet{Jafarzadeh_etal_2019} above that superpenumbra/penumbra comprises the vast majority of the plage pixels in that work. To this we add that the regions of plage used in \citet{Bastian_etal_2018} are only partially included within the FOV in \citet{Jafarzadeh_etal_2019} (i.e., further reducing the similarity of plage regions selected for the statistical studies in these two works). All these factors render the comparison with \citet{Bastian_etal_2018} a difficult task. To keep this discussion focused we moved some details to the APPENDIX.

Contrary to \citet{Bastian_etal_2018} who calculated the C.C. between \emph{ALMA}/Band6 $T_b$ and the mean $T_{\rm rad}$ from \ion{Mg}{2} h2v and h2r, the C.C. values in \citet{Jafarzadeh_etal_2019} were calculated separately for each individual line feature of \ion{Mg}{2} $T_{\rm rad}$ with \emph{ALMA}/Band6 $T_b$; namely: for \emph{ALMA}/Band6 $T_b$ vs. $T_{\rm rad}$ \ion{Mg}{2} k2v (and k2r) the C.C. was 0.73 (and 0.80), and 0.68 (0.78) for \ion{Mg}{2} h2v (h2r). 
Comparatively, the \emph{ALMA}/Band6 with \ion{Mg}{2} k2v (and k2r) was 48\% (and 49\%) higher than those calculated from our dataset of plage with line fittting; as for the \ion{Mg}{2} h2v (h2r), that was 39\% (42\%) higher than our values, respectively. Reconciling all the differences we mentioned between the work of \citet{Jafarzadeh_etal_2019} and our study and also considering the fact that both studies focused on pixels with the least time-difference possible, the discrepancies between our scatter plots and theirs can be understood due to their inclusion of pores or superpenumbra pixels in their statistics. If we do the same experiment as we did earlier in our comparison to \citet{Bastian_etal_2018} and we include the periphery in our calculations, then the C.C. reported in \citet{Jafarzadeh_etal_2019} for \ion{Mg}{2} k2v (and k2r) is only 7\% (and 14\%) higher than those calculated from our dataset; and for the \ion{Mg}{2} h2v (and h2r), that is only 2\% (and 11\%) higher than ours, respectively. In Figure 3 and Figure 5e of \citet{Jafarzadeh_etal_2019} the \emph{ALMA} and \emph{IRIS} maps and the map showing the mask used for plage is provided. By consulting the time-difference map between the \emph{ALMA} and \emph{IRIS} pixels shown in Figure 4f therein, we can identify exactly which parts of the plage were used in the scatter plots of Figure 13 therein. There appear to be several locations with radial dark ``lanes'' or ``streaks'' of low $T_{\rm rad}$ and $T_b$ around the superpenumbra/penumbra, and the majority of these pixels were used in the plage scatter plots of \citet{Jafarzadeh_etal_2019}. Therefore, it is possible that the similarity of the results for plage reported by \citet{Jafarzadeh_etal_2019} and \citet{Bastian_etal_2018} and their discrepancy with our results may have the same origin: the difference in the criteria used in distinguishing regions of plage from other neighboring regions on the Sun.

\section{Summary \& Conclusions}\label{conclusion}

In this work we focused on addressing the nature and the dynamics of chromospheric/transition region structures found in plage, namely, fibrils, jet-like features (\bf Type-II spicules are covered in a companion paper, Chintzoglou et al 2020a\rm) and traveling shocks using high time-cadence and high spatial resolution data from the \emph{ALMA} and \emph{IRIS} observatories. 
We employed a 2.5D numerical simulation (Bifrost model) of a plage region considering ambipolar diffusion in non-equilibrium ionization conditions. We produced synthetic observables to compare the model with our observations from \emph{ALMA}/Band6 and \emph{IRIS}. Last, we performed a first-cut study on the heating of the chromosphere above plage by measuring the brightness temperature modulation due to passing shocks with \emph{ALMA}/Band6. 

\begin{figure*}
	\includegraphics[width=\linewidth]{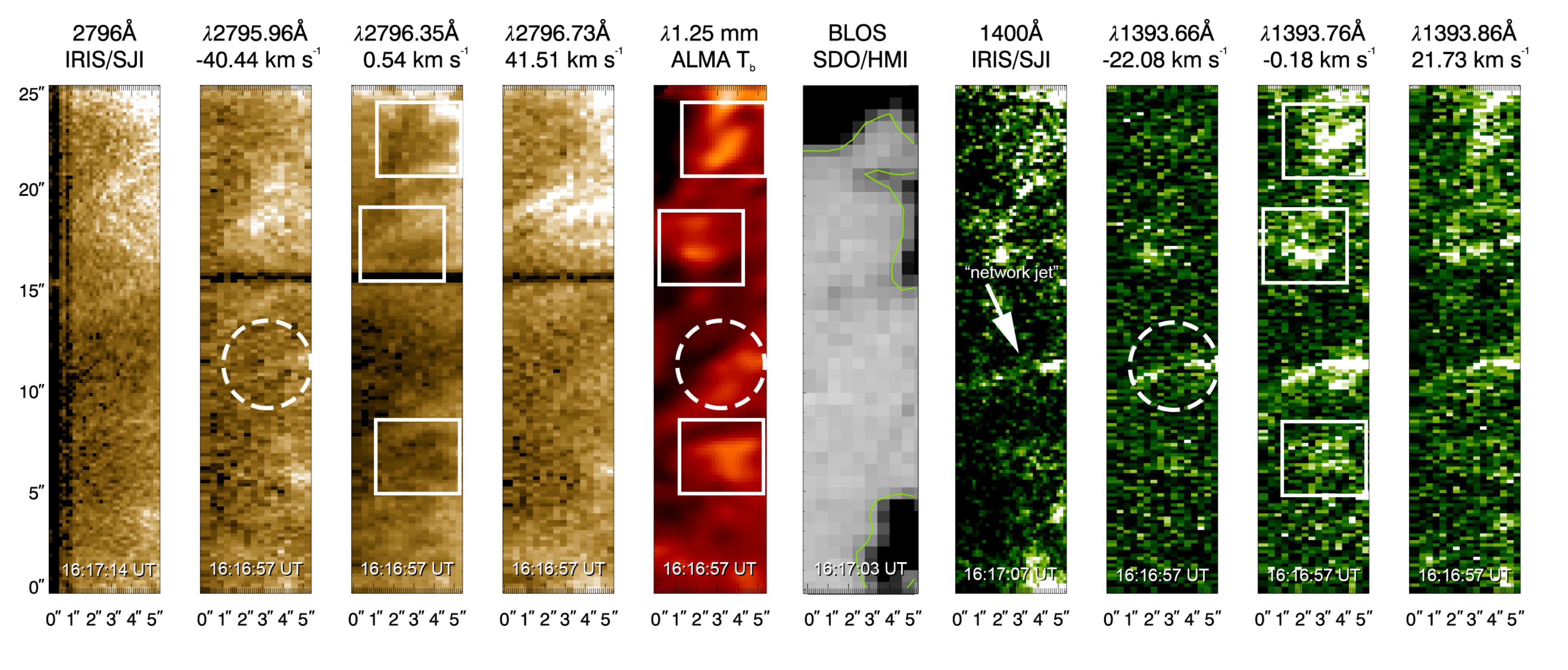}
	\caption{Inter-comparison of \emph{IRIS} and \emph{ALMA}/Band6 observations at an intermediate time in our observing window. Raster scans are shown at selected wavelength positions in \ion{Mg}{2} (left panels) and \ion{Si}{4} (right panels) showing the clear appearance of the rapidly evolving structure in \ion{Mg}{2} and \ion{Si}{4} (dashed circles). Boxed areas denote locations where intensity features in the \ion{Mg}{2} maps appear anti-correlated to those in \ion{Si}{4} and \emph{ALMA}/Band6 maps. In the panels with the \emph{SDO}/HMI magnetogram (scaling clipped at $\pm$250\,G) we overplot an isocontour of $\pm$100\,G. The FOV is thus split into two strongly magnetized areas (north and south of the FOV) separated by a weakly magnetized area (in the middle). \bf An animated version of this figure can be found in the online version of the journal. \rm}\label{FIG_2}
\end{figure*}
\begin{figure*}
        \includegraphics[width=5.5in]{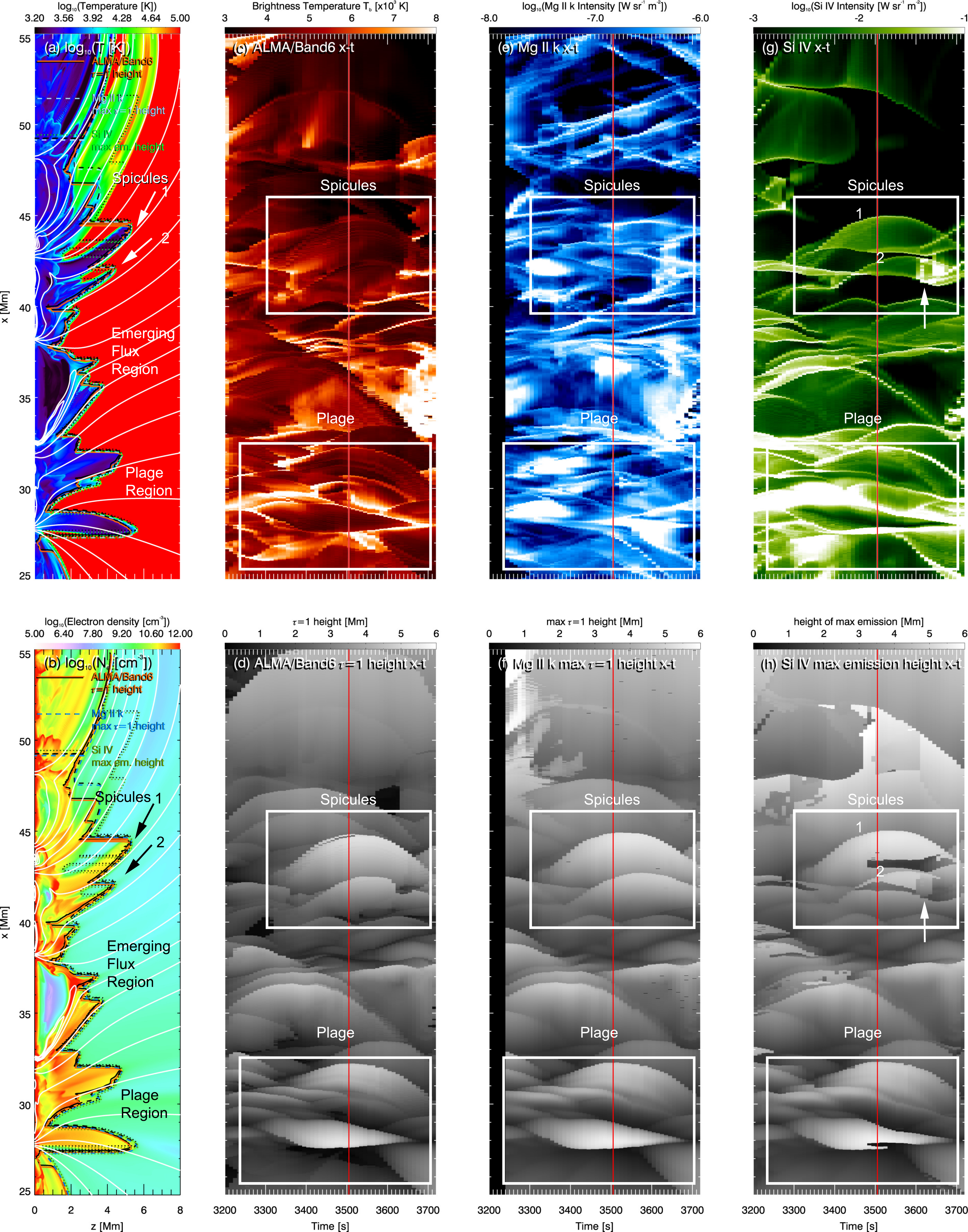}
	\caption{Synthetic $x-t$ plots from the Bifrost simulation. Left column panels show a snapshot from the simulation at $t$=3,500\,s for (a) $log_{10}T$, and (b) $log_{10} N_e$. Upper panels (c,e,g) show $x-t$ in \emph{ALMA}/Band6 T$_b$ (in red), \ion{Mg}{2} k (in blue), and \ion{Si}{4} (in green) synthesized intensities, respectively. Bottom panels (d, f, h) present $x-t$ plots for the geometric height where $\tau=1$ for \emph{ALMA}/Band6, maximum $\tau=1$ height for the wavelength-integrated \ion{Mg}{2} k, and the height of maximum emission for \ion{Si}{4}. Boxed areas denote the regions of spicules (best seen in \ion{Si}{4}) and plage. With a red line in the $x-t$ plots we mark the time shown in panels (a) and (b).	\bf An animated version of this figure can be found in the online version of the journal. \rm
	}\label{FIG_4}
\end{figure*}
\begin{figure*}
	\includegraphics[width=4.6in]{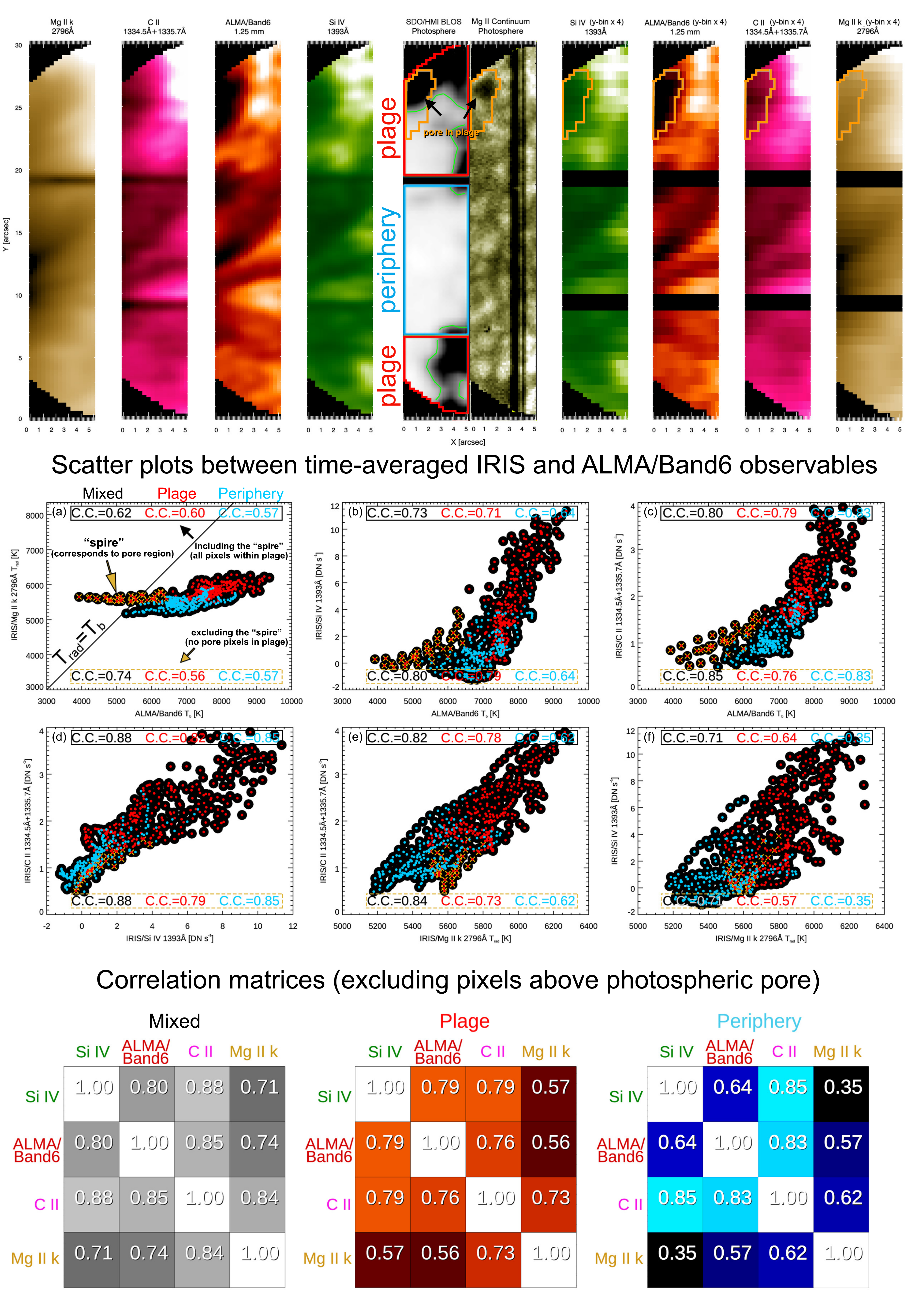}
	\caption{Top panels: time-averages of \ion{Mg}{2} k ($T_{\rm rad}$), \ion{C}{2}, and \ion{Si}{4} from \emph{IRIS}, and \emph{ALMA}/Band6 ($T_b$) together with their 4-pixel binned version along slit (y-axis). In the middle we show the time-averaged LOS magnetogram (green contour at $\pm$100\,G) with the subregions considered in the correlation plots (red/plage, blue/periphery of plage) and a \ion{Mg}{2} 2832\,\AA\ SJI map showing the presence of a photospheric pore in the plage region. Orange contours show pixel area determined as related to the pore, which is excluded in our analysis. Middle panels (a)-(f): Scatter plots for each combination of the 4-pix binned average series. Correlation coefficients (C.C.) are given for (i) points inside the plage area (shown in red), (ii) points inside the area containing the periphery of plage (blue), and (iii) for the entire FOV (black) with and without the pixels in the pore region (top  and bottom group of C.C. values, respectively). Note the significantly high correlation between \emph{ALMA}/Band6 and \ion{C}{2} and \ion{Si}{4}. Bottom panels: C.C. values excluding the pore pixels organized in correlation matrices. See text for detailed discussion.} \label{FIG_CII}
\end{figure*}
\begin{figure*}
	\includegraphics[width=\linewidth]{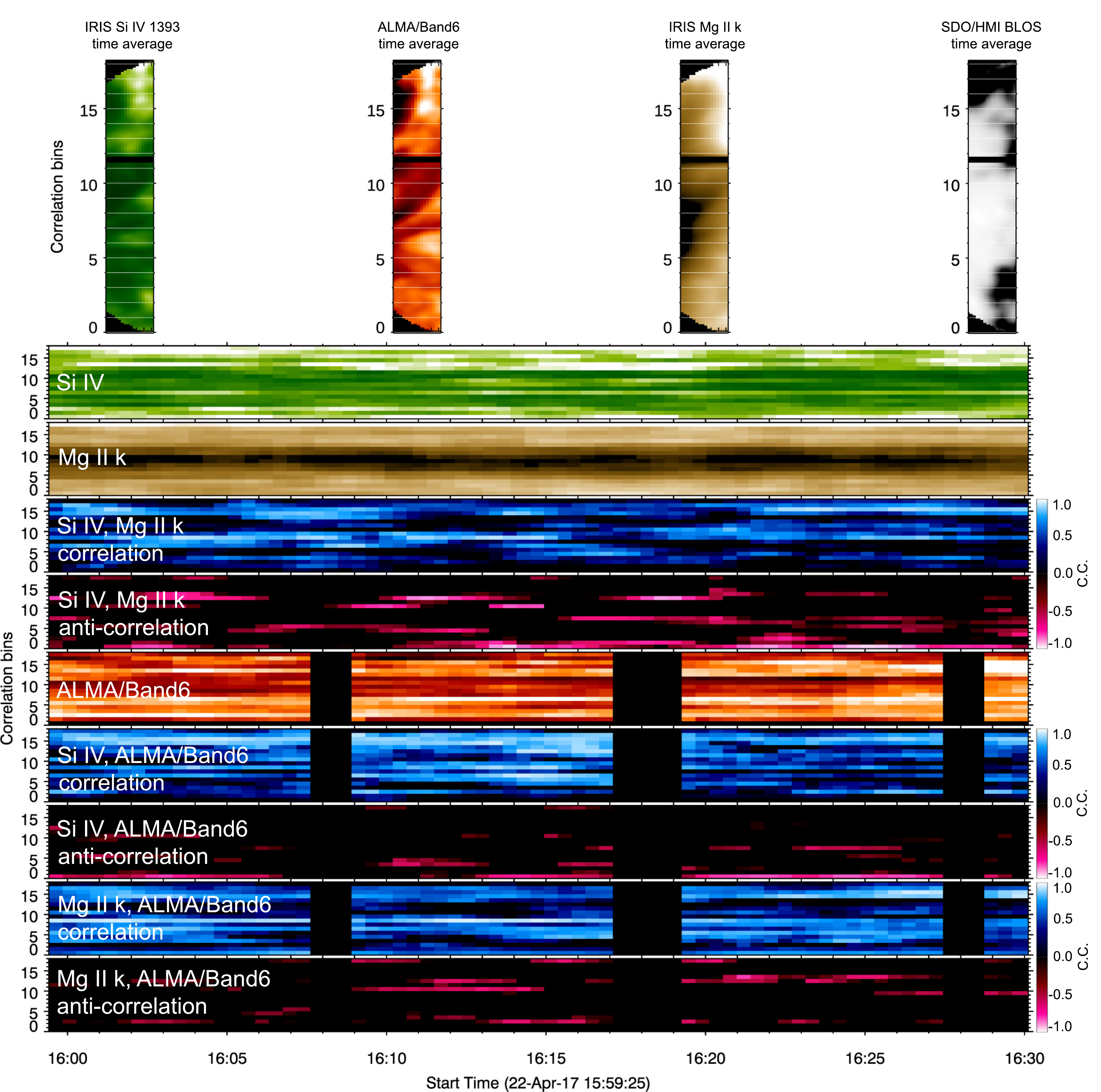}
	\caption{Time evolution of correlation and anti-correlation between \ion{Mg}{2}, \emph{ALMA}/Band6 and \ion{Si}{4}. The common FOV (covering the center of \emph{ALMA}/Band6 FOV) is segmented into 18 sub regions (correlation bins) covering both magnetic and non-magnetic regions (illustrated with time-averaged maps; top panels). Note that the \emph{IRIS} image raster series has been degraded with the \emph{ALMA}/Band6 beam size and position angle at each individual frame. Bottom panels: For illustration purposes we show the bin-averaged $x-t$ plot for \ion{Mg}{2} k, \emph{ALMA}/Band6 and \ion{Si}{4} 1393\,\AA. Note the clear distinction between areas of moderate-to-high correlation and anti-correlation between \ion{Mg}{2} k, \ion{Si}{4} 1393\,\AA. Note the very high degree of correlation between \ion{Si}{4} 1393\,\AA\ and \emph{ALMA}/Band6 across all correlation bins, and the more sporadic distribution of correlation for \ion{Mg}{2} k.
	} \label{FIG_RASTER_CORREL_STACK}
\end{figure*}
\begin{figure*}
	\includegraphics[width=\linewidth]{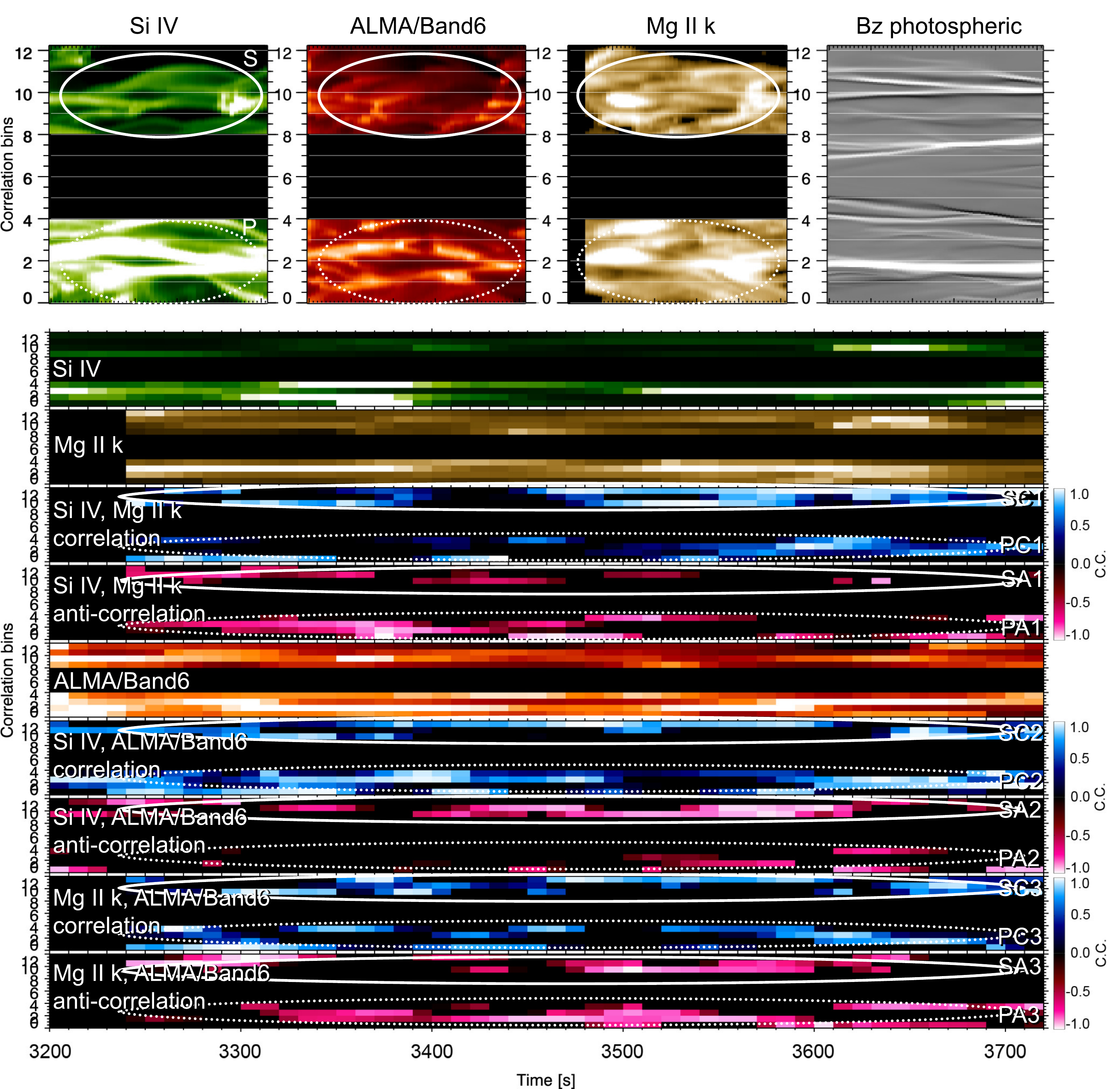}
	\caption{Similar as in Figure~\ref{FIG_RASTER_CORREL_STACK} with analysis method applied on the synthetic observables from the simulation. The spatial resolution in the \emph{IRIS}-\emph{ALMA} synthetic observables has been degraded accordingly to match the resolution of each observatory and then degraded \emph{IRIS} to match \emph{ALMA} (with an average \emph{ALMA}/Band6 beam size of 0$\farcs$8 along the y-direction;  x-direction is the time). The correlation bins have similar physical width across the y-direction as with previous Figure~\ref{FIG_RASTER_CORREL_STACK}. The emerging flux region is masked out (blank space) virtually isolating a region of spicules (top region) from a plage region (bottom region). Note the significant spatial extent of persistent anti-correlation between synthetic \ion{Mg}{2} k and \emph{ALMA}/Band6 observables.
	} \label{FIG_SIM_CORREL_STACK}
\end{figure*}
\begin{figure*}
	\includegraphics[width=\linewidth]{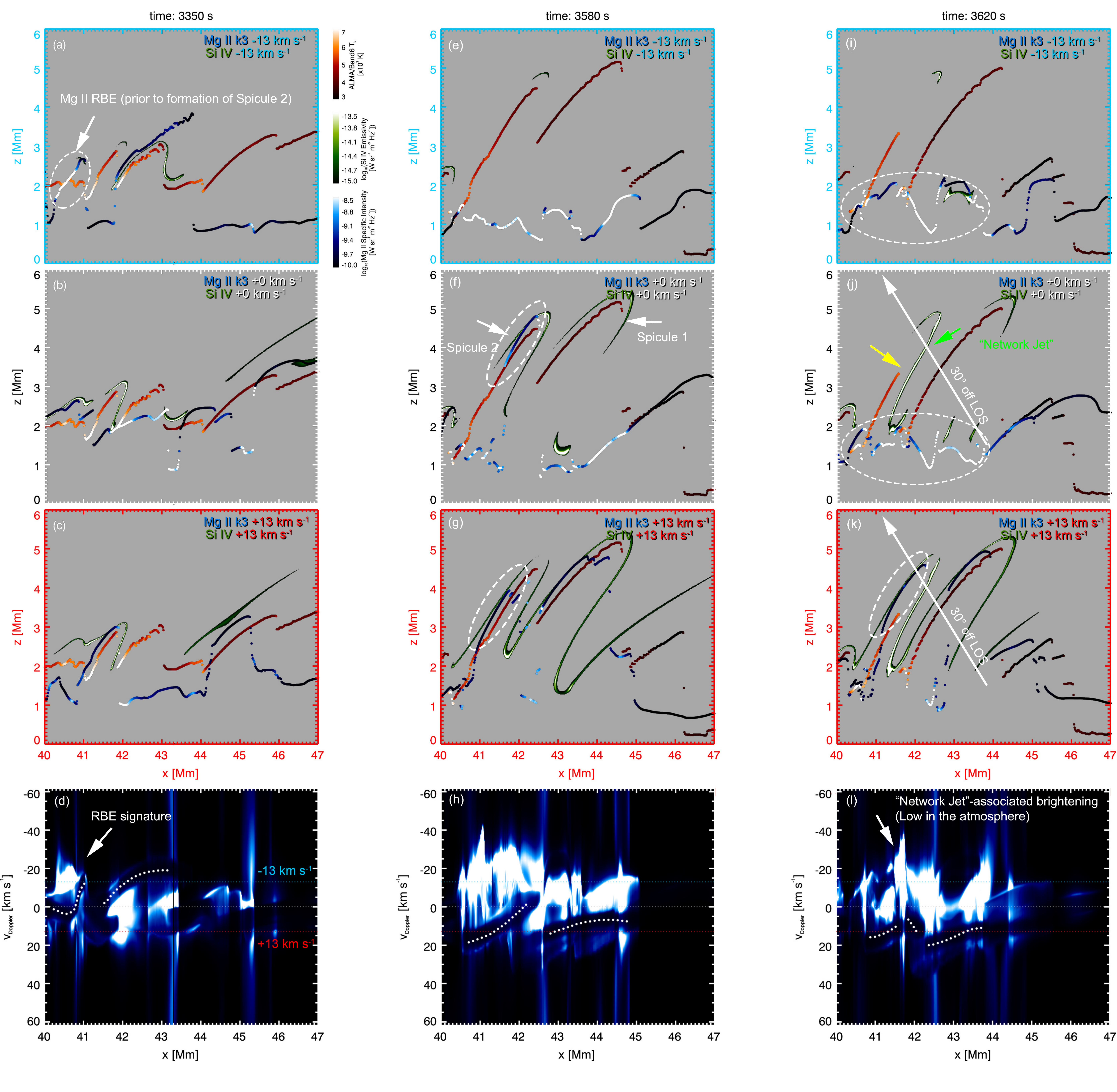}
	\caption{Maps from the simulation for the region with the spicules, showing the spatial distribution of \emph{ALMA}/Band6 emission from the spicules (at the height of $\tau=1$), along with \ion{Si}{4} emissivity and the \ion{Mg}{2} k specific intensity (at the height of $\tau=1$) at three different wavelength positions (corresponding to -13, 0, and +13 km s$^{-1}$) and at three different times (one per column). Note that at all times (at least until the brightening of the spicule at 3,620\,s, yellow arrow) \emph{ALMA}/Band6 follows closely the parts of the spicule emitting in \ion{Si}{4}. At 3,620\,s, the brightening occurs when the spicular mass is receding back to the surface, showing significant emission at redshifts (+13 km s$^{-1}$). Bottom row shows the \ion{Mg}{2} $\lambda-x$ plot with colored dotted lines denoting the wavelength positions shown above. The dark band seen in the spectra is a well-developed k3 component in \ion{Mg}{2} k (parts of it traced by a white dotted line). See text for discussion. \bf An animated version of this figure can be found in the online version of the journal. \rm
	} \label{FIG_SIM_EMISSIV_SPICULES}
\end{figure*}
\begin{figure*}
	\includegraphics[width=\linewidth]{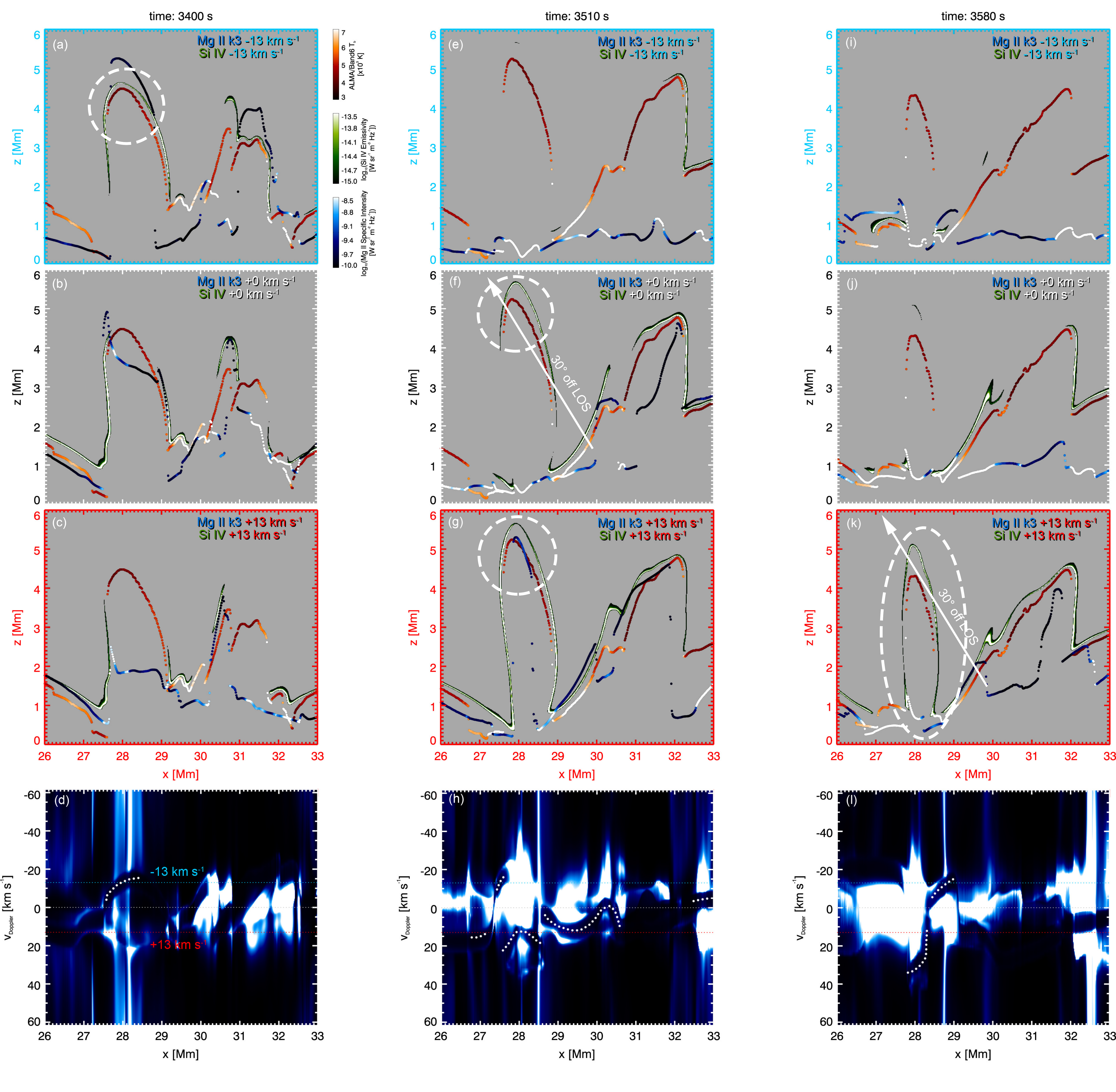}
	\caption{Maps from the simulation for the plage region, showing the spatial distribution of \emph{ALMA}/Band6 emission and intensity/emissivity from \emph{IRIS} observables at three different wavelength positions (corresponding to -13, 0, and +13 km s$^{-1}$) and at three different times (one per column, as in previous Figure~\ref{FIG_SIM_EMISSIV_SPICULES}). Note that the location of \emph{ALMA}/Band6 emission follows the locations of emissivity in \ion{Si}{4} more closely than \ion{Mg}{2} k intensity, which is consistent with the high degree of correlation of \emph{ALMA}/Band6 with \ion{Si}{4} seen in plage regions both in the simulation and in the observations. As in the previous figure, the bottom plots show the  \ion{Mg}{2} $\lambda-x$ plot with colored dotted lines denoting the wavelength positions shown above. The dark band seen in the spectra is a well-developed k3 component in \ion{Mg}{2} k (traced in part by a white dotted line). See text for discussion.
	} \label{FIG_SIM_EMISSIV_PLAGE}
\end{figure*}
\begin{figure*}
	\includegraphics[width=5in]{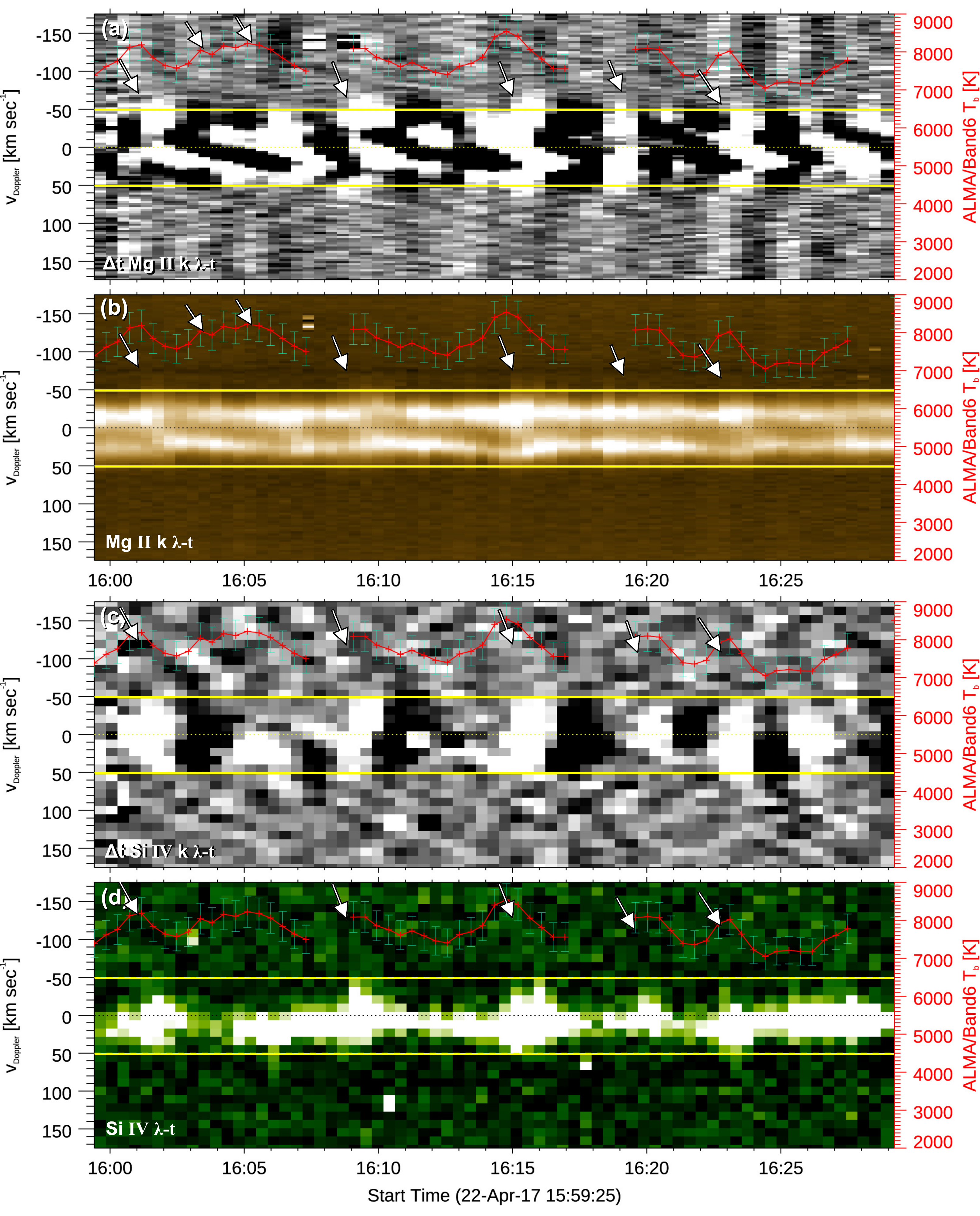}
	\caption{Top panels: $\lambda$-t plots for \ion{Mg}{2} k (b) and its time derivative (a) for the selected 1$\arcsec\times$1$\arcsec$ area above a plage region with recurrent shocks. Bottom panels: \ion{Si}{4} and (d) its corresponding time derivative (c). The $\lambda$-t panels are unsharp-masked to improve contrast in the presentation of the figure. In all panels we overplot the \emph{ALMA}/Band6 T$_b$. The error bars correspond to 5\% uncertainty in $T_b$ values. The rest wavelength position is plotted with a dotted line in all panels. Note the correlation of increases in T$_b$ and blue shifts suggesting chromospheric heating due to the passing of shocks in \ion{Si}{4} and \ion{Mg}{2} (pointed by arrows).}\label{FIG_8}
\end{figure*}

Below we summarize our findings:

\begin{enumerate}
	
\item We report a very high degree of similarity for features seen in plage between \emph{ALMA}/Band6 and \emph{IRIS}/SJI 1400\,\AA\ and \ion{Si}{4} 1393\,\AA\ rasters (Figure~\ref{FIG_2}). 
We conclude that \emph{ALMA}/Band6 is sensitive to the cool plasma at the highest parts of the spicules and dynamic fibrils (i.e., in plage), which is next (because of the locally large temperature gradients) to plasma that emits in \ion{Si}{4} ($T\approx$80,000K, transition region temperatures; this result also provides support to the work of \citealt{Rutten_2017}).

\item We present observations showing anti-correlation between intensity features seen in \ion{Si}{4} and \ion{Mg}{2} \emph{IRIS} rasters (intensity depressions \ion{Mg}{2} k highlighted in Figure~\ref{FIG_2}) 
We conclude that the apparent anti-correlation or lack of correlation 
has its origin in \ion{Mg}{2} opacity effects in plage structures. Strong absorption is the reason behind the low \ion{Mg}{2} intensities emerging from greater geometric heights in the locations of spicules (Figure~\ref{FIG_SIM_EMISSIV_SPICULES}). 

\item For plage we report a low linear correlation coefficient (0.49$\leq$C.C.$\leq$0.55) for \emph{ALMA}/Band6 $T_b$ with \ion{Mg}{2} for any of the k2v/k2r/h2v/h2r line features and a maximum of 0.56 for wavelength-integrated \ion{Mg}{2} k that contains the k2v, k3, k2r line features. 
Our results are quantitatively in contrast with previous works (e.g., \citealt{Bastian_etal_2018}, and \citealt{Jafarzadeh_etal_2019}).
We also determined that by including quieter areas in the plage sample, i.e., considering a mixed area of plage and periphery of plage as ``plage'', it greatly increased the C.C. between \emph{ALMA}/Band6 $T_b$ with $T_{\rm rad}$ from \ion{Mg}{2} k and h, producing values as high as those in \citet{Bastian_etal_2018} and \citet{Jafarzadeh_etal_2019}. 
We thus caution on the different criteria employed for defining plage regions, which may skew quantitative studies of correlations between different observables. 

\item The definition of the spatial extent of plage is not formally well-defined in the previous literature. Here, our empirical approach focuses on isolating plage from its surroundings in a different manner from previous approaches (\S~\ref{plagemethod}). 
We also add that the C.C. is a very sensitive statistical measure. A small amount of outliers in a scatter plot can greatly affect the C.C. value. Thus given the high sensitivity of C.C. and its application in cases where plage cannot be robustly classified with conventional techniques (e.g. due to the presence of pores, or being too close to sunspot penumbras) we caution that the C.C. values between $T_{\rm rad}$ and $T_b$ in plage can be actually lower, i.e., as low as those we report in the present paper.

\item It has been previously reported \citep{Bastian_etal_2017, Jafarzadeh_etal_2019} that there is scatter between radiative temperatures from \ion{Mg}{2} and brightness temperatures from \emph{ALMA}/Band6, with a suggested cause of the scatter being the decoupling of the \ion{Mg}{2} source function from the local conditions. Our work demonstrates that the scatter is more significant than previously thought and highlights another reason behind its nature: both the observations 
and the simulation suggest that the formation height for \emph{ALMA}/Band6 emission is above that of \ion{Mg}{2} for most wavelengths along the \ion{Mg}{2} line, even though both \ion{Mg}{2} and \emph{ALMA}/Band6 are sensitive to a similar range of temperatures. This formation height difference can contribute to the large scatter. 
We caution that the model should not be taken as a perfect representation of the plage atmosphere. For example, the distribution of modeled brightness temperatures of \emph{ALMA}/Band6 is lower by 2,000\,K as compared to the observations.

\item We performed a thorough quantitative study on the similarities between time-averaged \ion{Si}{4}, \ion{C}{2} (wavelength-integrated intensities), \ion{Mg}{2} k (average $T_{\rm rad}$), and \emph{ALMA}/Band6 ($T_b$) maps (Figure~\ref{FIG_CII}). We report that the highest C.C. is obtained between any combination of the triad of \ion{Si}{4}, \ion{C}{2} and \emph{ALMA}/Band6, the lowest C.C. is obtained between \ion{Mg}{2} k - \ion{Si}{4} and \ion{Mg}{2} k - \emph{ALMA}/Band6. Additionally, \ion{C}{2} is found in moderate-to-high C.C. with all other observables. We conclude that this study provides evidence on the general tendency for the order of the formation heights of all these different obsevables with geometric height. That is, \ion{Mg}{2} k (wavelength-integrated) intensity ($T_{\rm rad}$) emerges from lower geometric heights in the plage atmosphere, with \ion{Si}{4} forming at the greatest heights (with a formation temperature T$\approx$80,000\,K, placing it in the transition region). The good agreement of \ion{C}{2} with all observables is due to having, on average, a formation height between that of \ion{Mg}{2} k and \ion{Si}{4}. This result seems consistent with the work of \citet{Rathore_etal_2015b} based on the analysis of a numerical simulation.

\item We present indications of heating by shocks propagating in the chromosphere with \emph{ALMA}/Band6 (beam size $\approx$0$\farcs$7$\times$0$\farcs$8). We found a repetitive increase-and-decrease of the local chromospheric plasma temperature above plage of order 10-20\% from a basal value of 7,500\,K (for comparison, the temperature for a location in the periphery of plage was found $\approx$ 5,500\,K), with a decay time back to the baseline of about 60-120\,s (Figure~\ref{FIG_8}). We find indications of a recurrence at around 120\,s. This may suggest that a specific location in plage may be pervaded by shocks coming from different directions and at different timings, leading to intensity (and $T_b$) modulations of shorter (or even irregular) periods. \citet{DaSilvaSantos_etal_2019} performed inversions of \emph{IRIS} with \emph{ALMA} and reported that emission from shocks in the \emph{ALMA}/Band6 plage may be coming from lower optical depths (i.e., higher geometric heights) as compared to weakly magnetized areas (e.g., plage periphery). This is consistent with our determination of the formation height for \emph{ALMA}/Band6, which seems to be just below that of \ion{Si}{4}.


\end{enumerate}

This work demonstrates the benefits of the synergy between \emph{ALMA} and \emph{IRIS} observations, which effectively expanded the diagnostic capabilities of each observatory, and also tested and provided constrains for advanced numerical simulations.

\acknowledgements

This paper makes use of the following \emph{ALMA} data: ADS/JAO.ALMA\#2016.1.00050.S. \emph{ALMA} is a partnership of ESO (representing its member states), NSF (USA) and NINS (Japan), together with NRC (Canada), MOST and ASIAA (Taiwan), and KASI (Republic of Korea), in cooperation with the Republic of Chile. The Joint \emph{ALMA} Observatory is operated by ESO, AUI/NRAO and NAOJ. We gratefully acknowledge support by NASA contract NNG09FA40C (IRIS). JMS is also supported by NASA grants NNX17AD33G, 80NSSC18K1285 and NSF grant AST1714955. VH is supported by NASA grant 80NSSC20K1272.
JdlCR is supported by grants from the Swedish Research Council (2015-03994), the Swedish National Space Board (128/15) and the Swedish Civil Contingencies Agency (MSB). This project has received funding from the European Research Council (ERC) under the European Union's Horizon 2020 research and innovation programme (SUNMAG, grant agreement 759548). 
MS, SJ and SW are supported by the SolarALMA project, which has received funding from the European Research Council (ERC) under the European Union’s Horizon 2020 research and innovation programme (grant agreement No. 682462), and by the Research Council of Norway through its Centres of Excellence scheme, project number 262622.
The simulations and \ion{Mg}{2} synthesis were ran on clusters from the Notur project, and the Pleiades cluster through the computing project s1061, s1630, and s2053 from the High End Computing (HEC) division of NASA. This research is also supported by the Research Council of Norway through 
its Centres of Excellence scheme, project number 262622, and through 
grants of computing time from the Programme for Supercomputing. 
\emph{IRIS} is a NASA small explorer mission developed and operated by LMSAL with mission operations executed at NASA Ames Research center and major contributions to downlink communications funded by ESA and the Norwegian Space Centre. HMI and AIA are instruments on board \emph{SDO}, a mission for NASA’s Living with a Star program.

\clearpage

\clearpage

\clearpage



\clearpage

\clearpage

\clearpage

\clearpage

%

\clearpage

\section{APPENDIX}\label{APPENDIX}
\appendix 

To improve the readability of the main text we host here part of the discussion found in \S~\ref{discrepancies} where we highlight the large discrepancies between our results (and our methodology) with previous studies (e.g., those presented in \citealt{Bastian_etal_2018, Jafarzadeh_etal_2019}). In particular, here we highlight the effects due to the inclusion (or insufficient exclusion) of pores in plage as a source of bias, and we also emphasize our unprecedented time-synchronization between rapid EUV rasters and rapid mm-emission imaging observations in comparison to that achieved by previous studies.

In contrast to \citet{Bastian_etal_2018}, \citet{Jafarzadeh_etal_2019} attempt to remove the pores within the plage by applying a mask derived from an HMI photospheric magnetogram with magnetic field values $\geq\pm $0.8\,kG. However, we note here that the small size of the pores in that plage cannot be fully accounted for by the simple application of a mask from a photospheric magnetogram, due to the natural expansion of fields as they reach chromosperic and transition region heights. Conversely, this makes any pore region mask produced from observations at the photosphere to contain only a part of the associated area higher up, making the safe extraction of pores from the immediate plage a challenging task. This is due to superposition effects and confusion along the line of sight of structures in the chromospheric data. The definition of pore regions within plage in \citet{Jafarzadeh_etal_2019} appears to include real plage signal. In Figures 6 and 7 of \citet{Jafarzadeh_etal_2019} we can see that the pore signal is included in the histograms done for other regions in the FOV, namely ``Penumbra and Pores'', ``Sunspot and Pores'', where the \ion{Mg}{2} k2v and k2r and h2v and h2r are clearly skewed to the higher $T_{\rm rad}$ with a sharp drop at 6,000\,K. However, the histograms for ``Umbra'' and ``Quiet Regions'' are clearly skewed to the low end of $T_{\rm rad}$ with a very gradual drop towards 6,000\,K. Thus, they share similarities with the histograms for ``Plage'', justifying our concerns for proper characterization of plage from non-plage regions. In our work here, while we are not studying the region above the pore, we are safely excluding it by applying a threshold on \emph{ALMA}/Band6 $T_b$; this allows to account for the more extended boundaries of structures as they expand from the photosphere to the chromosphere, leaving behind a clean map for plage (Figure~\ref{FIG_CII}). 

Also, we note that \citet{Jafarzadeh_etal_2019} distinguished the \emph{ALMA}/Band6 data into the four sub-bands and only used the one at $\lambda$=1.3\,mm, instead of taking the average of all sub-bands as in \citet{Bastian_etal_2018} and in our present work. Such averaging results to Band6 maps at $\lambda$=1.25\,mm. Finally, \citet{Jafarzadeh_etal_2019} strived to take into account the time-differences between the \emph{ALMA}/Band6 mosaic and the scanning time of the large \emph{IRIS} raster. Unfortunately, due to the nature of the mosaicking scanning process of these particular \emph{ALMA}/Band6 observations (complicated since it does not follow the scanning direction of the \emph{IRIS} raster, which takes non-trivial amount of time) the authors had only a limited amount of pixels with a minimal time-difference in their dataset. Thus, despite the attempts to match the time between pixels from \emph{ALMA} and \emph{IRIS}, an adequate amount of pixels for their statistics was obtained only after allowing for a variable matching of the sampling time, i.e., spanning 0.5-2\,min (Figure 4f in \citealt{Jafarzadeh_etal_2019}). In addition, the same work explores correlations between observables by limiting the time-differences to 0.5\,min at the expense of sample number. In comparison, \citet{Bastian_etal_2018} did not select pixels with such criteria, thus significant chromospheric evolution is not captured in that analysis. As we mentioned in the beginning of \S~\ref{discrepancies}, our sampling time synchronization between datasets is superb, i.e., $\pm$1\,s at worst and is consistent throughout the data series analyzed in our work. This is due to the fast \emph{IRIS} raster scanning times (26\,s) for this particular observation and also thanks to the rapid cadence (2\,s) of our \emph{ALMA}/Band6 observations.


\begin{thebibliography}{33}
	\expandafter\ifx\csname natexlab\endcsname\relax\def\natexlab#1{#1}\fi
	
	\bibitem[{{Bastian} {et~al.}(2017){Bastian}, {Chintzoglou}, {De Pontieu},
		{Shimojo}, {Schmit}, {Leenaarts}, \& {Loukitcheva}}]{Bastian_etal_2017}
	{Bastian}, T.~S., {Chintzoglou}, G., {De Pontieu}, B., {et~al.} 2017, \apjl,
	845, L19
	
	\bibitem[{{Bastian} {et~al.}(2018){Bastian}, {Chintzoglou}, {De Pontieu},
		{Shimojo}, {Schmit}, {Leenaarts}, \& {Loukitcheva}}]{Bastian_etal_2018}
	---. 2018, \apjl, 860, L16
	
	\bibitem[{{Carlsson} {et~al.}(2019){Carlsson}, {De Pontieu}, \&
		{Hansteen}}]{Carlsson_etal_2019}
	{Carlsson}, M., {De Pontieu}, B., \& {Hansteen}, V.~H. 2019, \araa, 57, 189
	
	\bibitem[{{Carlsson} {et~al.}(2015){Carlsson}, {Leenaarts}, \& {De
			Pontieu}}]{Carlsson_etal_2015}
	{Carlsson}, M., {Leenaarts}, J., \& {De Pontieu}, B. 2015, \apjl, 809, L30
	
	\bibitem[{{Carlsson} \& {Stein}(1997)}]{Carlsson_Stein_1997}
	{Carlsson}, M., \& {Stein}, R.~F. 1997, \apj, 481, 500
	
	\bibitem[{{Carlsson} \& {Stein}(2002)}]{Carlsson_Stein_2002}
	---. 2002, \apj, 572, 626
	
	\bibitem[{{Chintzoglou} {et~al.}(2018){Chintzoglou}, {De Pontieu},
		{Mart{\'\i}nez-Sykora}, {Pereira}, {Vourlidas}, \& {Tun
			Beltran}}]{Chintzoglou_etal_2018}
	{Chintzoglou}, G., {De Pontieu}, B., {Mart{\'\i}nez-Sykora}, J., {et~al.} 2018,
	\apj, 857, 73
	
	\bibitem[{{Chintzoglou} {et~al.}(2017){Chintzoglou}, {Vourlidas}, {Savcheva},
		{Tassev}, {Tun Beltran}, \& {Stenborg}}]{Chintzoglou_etal_2017}
	{Chintzoglou}, G., {Vourlidas}, A., {Savcheva}, A., {et~al.} 2017, \apj, 843,
	93
	
	\bibitem[{{Chintzoglou} {et~al.}(2019){Chintzoglou}, {Zhang}, {Cheung}, \&
		{Kazachenko}}]{Chintzoglou_etal_2019}
	{Chintzoglou}, G., {Zhang}, J., {Cheung}, M. C.~M., \& {Kazachenko}, M. 2019,
	\apj, 871, 67
	
	\bibitem[{{da Silva Santos} {et~al.}(2020){da Silva Santos}, {de la Cruz
			Rodr{\'\i}guez}, {Leenaarts}, {Chintzoglou}, {De Pontieu}, {Wedemeyer}, \&
		{Szydlarski}}]{DaSilvaSantos_etal_2019}
	{da Silva Santos}, J.~M., {de la Cruz Rodr{\'\i}guez}, J., {Leenaarts}, J.,
	{et~al.} 2020, \aap, 634, A56
	
	\bibitem[{{De Pontieu} {et~al.}(2007{\natexlab{a}}){De Pontieu}, {Hansteen},
		{Rouppe van der Voort}, {van Noort}, \& {Carlsson}}]{DePontieu_etal_2007a}
	{De Pontieu}, B., {Hansteen}, V.~H., {Rouppe van der Voort}, L., {van Noort},
	M., \& {Carlsson}, M. 2007{\natexlab{a}}, \apj, 655, 624
	
	\bibitem[{{De Pontieu} {et~al.}(2007{\natexlab{b}}){De Pontieu}, {McIntosh},
		{Carlsson}, {Hansteen}, {Tarbell}, {Schrijver}, {Title}, {Shine}, {Tsuneta},
		{Katsukawa}, {Ichimoto}, {Suematsu}, {Shimizu}, \&
		{Nagata}}]{DePontieu_etal_2007b}
	{De Pontieu}, B., {McIntosh}, S.~W., {Carlsson}, M., {et~al.}
	2007{\natexlab{b}}, Science, 318, 1574
	
	\bibitem[{{De Pontieu} {et~al.}(2014){De Pontieu}, {Title}, {Lemen}, {Kushner},
		{Akin}, {Allard}, {Berger}, {Boerner}, {Cheung}, {Chou}, {Drake}, {Duncan},
		{Freeland}, {Heyman}, {Hoffman}, {Hurlburt}, {Lindgren}, {Mathur}, {Rehse},
		{Sabolish}, {Seguin}, {Schrijver}, {Tarbell}, {W{\"u}lser}, {Wolfson},
		{Yanari}, {Mudge}, {Nguyen-Phuc}, {Timmons}, {van Bezooijen}, {Weingrod},
		{Brookner}, {Butcher}, {Dougherty}, {Eder}, {Knagenhjelm}, {Larsen},
		{Mansir}, {Phan}, {Boyle}, {Cheimets}, {DeLuca}, {Golub}, {Gates}, {Hertz},
		{McKillop}, {Park}, {Perry}, {Podgorski}, {Reeves}, {Saar}, {Testa}, {Tian},
		{Weber}, {Dunn}, {Eccles}, {Jaeggli}, {Kankelborg}, {Mashburn}, {Pust},
		{Springer}, {Carvalho}, {Kleint}, {Marmie}, {Mazmanian}, {Pereira}, {Sawyer},
		{Strong}, {Worden}, {Carlsson}, {Hansteen}, {Leenaarts}, {Wiesmann},
		{Aloise}, {Chu}, {Bush}, {Scherrer}, {Brekke}, {Martinez-Sykora}, {Lites},
		{McIntosh}, {Uitenbroek}, {Okamoto}, {Gummin}, {Auker}, {Jerram}, {Pool}, \&
		{Waltham}}]{DePontieu_etal_2014}
	{De Pontieu}, B., {Title}, A.~M., {Lemen}, J.~R., {et~al.} 2014, \solphys, 289,
	2733
	
	\bibitem[{{Deslandres}(1893)}]{Deslandres_1893}
	{Deslandres}, H. 1893, Knowledge: An Illustrated Magazine of Science, 16, 230
	
	\bibitem[{{Hansteen} {et~al.}(2006){Hansteen}, {De Pontieu}, {Rouppe van der
			Voort}, {van Noort}, \& {Carlsson}}]{Hansteen_etal_2006}
	{Hansteen}, V.~H., {De Pontieu}, B., {Rouppe van der Voort}, L., {van Noort},
	M., \& {Carlsson}, M. 2006, \apjl, 647, L73
	
	\bibitem[{{Harvey} \& {Harvey}(1973)}]{Harvey_Harvey_1973}
	{Harvey}, K., \& {Harvey}, J. 1973, \solphys, 28, 61
	
	\bibitem[{{Jafarzadeh} {et~al.}(2019){Jafarzadeh}, {Wedemeyer}, {Szydlarski},
		{De Pontieu}, {Rezaei}, \& {Carlsson}}]{Jafarzadeh_etal_2019}
	{Jafarzadeh}, S., {Wedemeyer}, S., {Szydlarski}, M., {et~al.} 2019, \aap, 622,
	A150
	
	\bibitem[{{Langangen} {et~al.}(2008){Langangen}, {De Pontieu}, {Carlsson},
		{Hansteen}, {Cauzzi}, \& {Reardon}}]{Langangen_etal_2008}
	{Langangen}, {\O}., {De Pontieu}, B., {Carlsson}, M., {et~al.} 2008, \apjl,
	679, L167
	
	\bibitem[{{Loukitcheva} {et~al.}(2015){Loukitcheva}, {Solanki}, {Carlsson}, \&
		{White}}]{Loukitcheva_etal_2015}
	{Loukitcheva}, M., {Solanki}, S.~K., {Carlsson}, M., \& {White}, S.~M. 2015,
	\aap, 575, A15
	
		\bibitem[{{Mart{\'\i}nez-Sykora}
		{et~al.}(2020{\natexlab{a}}){Mart{\'\i}nez-Sykora}, {De Pontieu}, {de la Cruz
			Rodriguez}, \& {Chintzoglou}}]{Martinez-Sykora_etal_2020b}
	{Mart{\'\i}nez-Sykora}, J., {De Pontieu}, B., {de la Cruz Rodriguez}, J., \&
	{Chintzoglou}, G. 2020{\natexlab{a}}, \apjl, 891, L8
	
	\bibitem[{{Mart{\'\i}nez-Sykora} {et~al.}(2020){Mart{\'\i}nez-Sykora},
		{Leenaarts}, {De Pontieu}, {N{\'o}brega-Siverio}, {Hansteen}, {Carlsson}, \&
		{Szydlarski}}]{Martinez-Sykora_etal_2020a}
	{Mart{\'\i}nez-Sykora}, J., {Leenaarts}, J., {De Pontieu}, B., {et~al.} 2020{\natexlab{b}},
	\apj, 889, 95
	
	\bibitem[{{Pesnell} {et~al.}(2012){Pesnell}, {Thompson}, \&
		{Chamberlin}}]{Pesnell_etal_2012}
	{Pesnell}, W.~D., {Thompson}, B.~J., \& {Chamberlin}, P.~C. 2012, \solphys,
	275, 3
	
	\bibitem[{{Rathore} {et~al.}(2015{\natexlab{a}}){Rathore}, {Carlsson},
		{Leenaarts}, \& {De Pontieu}}]{Rathore_etal_2015a}
	{Rathore}, B., {Carlsson}, M., {Leenaarts}, J., \& {De Pontieu}, B.
	2015{\natexlab{a}}, \apj, 811, 81
	
	\bibitem[{{Rathore} {et~al.}(2015{\natexlab{b}}){Rathore}, {Pereira},
		{Carlsson}, \& {De Pontieu}}]{Rathore_etal_2015b}
	{Rathore}, B., {Pereira}, T. M.~D., {Carlsson}, M., \& {De Pontieu}, B.
	2015{\natexlab{b}}, \apj, 814, 70
	
	\bibitem[{{Rutten}(2017)}]{Rutten_2017}
	{Rutten}, R.~J. 2017, \aap, 598, A89
	
	\bibitem[{{Schmit} {et~al.}(2015){Schmit}, {Bryans}, {De Pontieu}, {McIntosh},
		{Leenaarts}, \& {Carlsson}}]{Schmit_etal_2015}
	{Schmit}, D., {Bryans}, P., {De Pontieu}, B., {et~al.} 2015, \apj, 811, 127
	
	\bibitem[{{Secchi}(1877)}]{Secchi_1877}
	{Secchi}, A. 1877, {Le Soleil. Seconde partie. Structure du Soleil (suite) --
		Activite ext\'erieure} (Gauthier Villars, Paris)
	
	\bibitem[{{Skogsrud} {et~al.}(2016){Skogsrud}, {Rouppe van der Voort}, \& {De
			Pontieu}}]{Skogsrud_etal_2016}
	{Skogsrud}, H., {Rouppe van der Voort}, L., \& {De Pontieu}, B. 2016, \apj,
	817, 124
	
	\bibitem[{{Wedemeyer} {et~al.}(2020){Wedemeyer}, {Szydlarski}, {Jafarzadeh},
		{Eklund}, {Guevara Gomez}, {Bastian}, {Fleck}, {de la Cruz Rodriguez},
		{Rodger}, \& {Carlsson}}]{Wedemeyer_etal_2020}
	{Wedemeyer}, S., {Szydlarski}, M., {Jafarzadeh}, S., {et~al.} 2020, \aap, 635,
	A71
	
	\bibitem[{{Wedemeyer-B{\"o}hm} {et~al.}(2007){Wedemeyer-B{\"o}hm}, {Ludwig},
		{Steffen}, {Leenaarts}, \& {Freytag}}]{Wedemeyer_etal_2007}
	{Wedemeyer-B{\"o}hm}, S., {Ludwig}, H.~G., {Steffen}, M., {Leenaarts}, J., \&
	{Freytag}, B. 2007, \aap, 471, 977
	
	\bibitem[{{Wootten} \& {Thompson}(2009)}]{Wootten_Thompson_2009}
	{Wootten}, A., \& {Thompson}, A.~R. 2009, IEEE Proceedings, 97, 1463
	
	\bibitem[{{Yurchyshyn} {et~al.}(2001){Yurchyshyn}, {Wang}, \&
		{Goode}}]{Yurchyshyn_etal_2001a}
	{Yurchyshyn}, V.~B., {Wang}, H., \& {Goode}, P.~R. 2001, \apj, 550, 470
	
	\bibitem[{{Zhang} {et~al.}(2003){Zhang}, {Solanki}, \&
		{Wang}}]{Zhang_Jun_etal_2003}
	{Zhang}, J., {Solanki}, S.~K., \& {Wang}, J. 2003, \aap, 399, 755
	
\end{thebibliography}
\end{document}